%% file: ms_emulateapj.tex
\documentclass[twocolumn]{aastex631}

\usepackage{xspace}
\usepackage{graphicx}
\usepackage{natbib}
\def\errtwo#1#2#3{$#1^{+#2}_{-#3}$}

\newcommand\msun{\mathrm{M_\odot}}

\newcommand\nh{N_\mathrm{H}}
\newcommand\cmt{\mathrm{cm^{-2}}}

\newcommand\arf{{\tt arf}\xspace}

\newcommand\chandra{\textsl{Chandra}\xspace}
\newcommand\cmtwo{\mathrm{cm^2}}
\newcommand\constant{\texttt{constant}\xspace}

\newcommand\cyg{{\mbox Cyg~X-1}\xspace}
\newcommand\dbh{D_\mathrm{BH}}
\newcommand\diskpn{\texttt{diskpn}\xspace}

\newcommand\emcee{\texttt{emcee}\xspace}

\newcommand\eqpair{\texttt{eqpair}\xspace}

\newcommand\fc{f_\mathrm{c}}
\newcommand\fpma{{FPMA}\xspace}
\newcommand\fpmb{{FPMB}\xspace}
\newcommand\fu{{4U~1957+11}\xspace}
\newcommand\gabs{\texttt{gabs}\xspace}
\newcommand\gaia{\textsl{Gaia}\xspace}

\newcommand\gx{{GX339$-$4}\xspace}
\newcommand\heasoft{\texttt{HEASOFT v6.29c}\xspace}

\newcommand\isis{\texttt{ISIS}\xspace}
\newcommand\kerrbb{\texttt{kerrbb}\xspace}
\newcommand\ktdisk{kT_\mathrm{disk}}
\newcommand\lc{l_\mathrm{c}}
\newcommand\ledd{L_\mathrm{Edd}}
\newcommand\lerat{L/L_\mathrm{Edd}}

\newcommand\lmcone{{LMC~X-1}\xspace}
\newcommand\lmcthree{{LMC~X-3}\xspace}
\newcommand\mbh{M_\mathrm{BH}}
\newcommand\maxi{{MAXI}\xspace}

\newcommand\neqp{N_\mathrm{eqp}}
\newcommand\nicer{\textsl{NICER}\xspace}
\newcommand\nicerarf{\texttt{nicerarf}\xspace}
\newcommand\nicerrmf{\texttt{nicerrmf}\xspace}
\newcommand\nustar{\textsl{NuSTAR}\xspace}
\newcommand\pca{{PCA}\xspace}

\newcommand\polykerrbb{\texttt{polykerrbb}\xspace}
\newcommand\powerlaw{\texttt{powerlaw}\xspace}
\newcommand\rd{R_\mathrm{d}}
\newcommand\relxillCp{\texttt{relxillCp}\xspace}
\newcommand\rin{R_\mathrm{in}}
\newcommand\rmf{\texttt{rmf}\xspace}

\newcommand\rossi{\textsl{Rossi X-ray Timing Explorer}\xspace}
\newcommand\rxte{\textsl{RXTE}\xspace}

\newcommand\suzaku{\textsl{Suzaku}\xspace}

\newcommand\swift{\textsl{Swift}\xspace}
\newcommand\tbabs{\texttt{tbabs}\xspace}
\newcommand\thcomp{\texttt{thcomp}\xspace}
\newcommand\xdust{x_\mathrm{dust}}

\newcommand\xscat{\texttt{xscat}\xspace}

\newcommand\taues{\tau_\mathrm{es}}
\newcommand\tc{T_\mathrm{c}}
\newcommand\teff{T_\mathrm{eff}}

\newcommand\aproxgt{\mathrel{%
     \rlap{\raise 0.511ex \hbox{$>$}}{\lower 0.511ex \hbox{$\sim$}}}}
\newcommand\aproxlt{\mathrel{%
     \rlap{\raise 0.511ex \hbox{$<$}}{\lower 0.511ex \hbox{$\sim$}}}}

\received{2022 November}
\accepted{2022 December}

\submitjournal{\apj}

\shorttitle{\nicer/\nustar Characterization of \fu} 
\shortauthors{Barillier et al.}

\begin{document}

\title{\nicer/\nustar Characterization of \fu: A Near Maximally
  Spinning Black Hole Potentially in the Mass Gap}

% Alphabetical Author List for Now
%
\author[0000-0001-9107-6542]{Erin Barillier}
\affiliation{Physics Dept., CB 1105, Washington University, One
  Brookings Drive, St. Louis, MO 63130-4899}
\affiliation{Dept.\ of Physics, Randall Lab, University of Michigan, 450 Church Street
  Ann Arbor, MI 48109-1040 }

\author[0000-0003-2538-0188]{Victoria Grinberg} 
\affiliation{European Space Agency (ESA), European Space Research and
  Technology Centre (ESTEC), Keplerlaan 1, 2201 AZ Noordwijk, The
  Netherlands}
\author[0000-0002-6288-4791]{David Horn}
\affiliation{Dr.\ Karl Remeis-Observatory \& ECAP, Friedrich-Alexander-Universit\"at Erlangen-N\"urnberg,
  Sternwartstr.~7, 96049 Bamberg, Germany}
\author[0000-0001-6923-1315]{Michael A. Nowak}
\affiliation{Physics Dept., CB 1105, Washington University, One
  Brookings Drive, St.\ Louis, MO 63130-4899}
\author[0000-0003-4815-0481]{Ronald A. Remillard} 
\affiliation{MIT
  Kavli Institute for Astrophysics and Space Research, Cambridge, MA
  02139, USA}
\author[0000-0002-5872-6061]{James F. Steiner}
\affiliation{Center for Astrophysics | Harvard \& Smithsonian, 60 Garden 
  Street, Cambridge, MA 02138, USA}
\author[0000-0001-5819-3552]{Dominic J. Walton}
\affiliation{Centre for Astrophysics Research, University of
  Hertfordshire, College Lane, Hatfield AL10 9AB, UK}
\affiliation{Institute
  of Astronomy, University of Cambridge, Madingley Road, Cambridge CB3
  0HA, UK}
\author[0000-0003-2065-5410]{J\"orn Wilms}
\affiliation{Dr.\ Karl Remeis-Observatory \& ECAP,
  Friedric-Alexander-Universit\"at Erlangen-N\"urnberg,
  Sternwartstr.~7, 96049 Bamberg, Germany}
\email{erinbari@umich.edu, victoria.grinberg@esa.int,
  david.horn@fau.de,
  mnowak@physics.wustl.edu,   
  ronrem4@gmail.com,
  james.steiner@cfa.harvard.edu, walton@ast.cam.ac.uk,
  joern.wilms@sternwarte.uni-erlangen.de}

\begin{abstract}
\fu is a black hole candidate system that has been in a soft X-ray
spectral state since its discovery.  We present analyses of recent
joint \nicer and \nustar spectra, which are extremely well-described
by a highly inclined disk accreting into a near maximally spinning
black hole. Owing to the broad X-ray coverage of \nustar the fitted
spin and inclination are strongly constrained for our hypothesized
disk models.  The faintest spectra are observed out to 20\,keV, even
though their hard tail components are almost absent when described
with a simple corona.  The hard tail increases with luminosity, but
shows clear two track behavior with one track having appreciably
stronger tails.  The disk spectrum color-correction factor is
anti-correlated with the strength of the hard tail (e.g., as measured
by the Compton $y$ parameter). Although the spin and inclination
parameters are strongly constrained for our chosen model, the mass and
distance are degenerate parameters.  We use our spectral fits, along
with a theoretical prior on color-correction, an observational prior
on likely fractional Eddington luminosity, and an observational prior
on distance obtained from \gaia studies, to present mass and distance
contours for this system.  The most likely parameters, given our
presumed disk model, suggest a 4.6\,$\msun$ black hole at 7.8\,kpc
observed at luminosities ranging from $\approx 1.7\%$--9\% of
Eddington.  This would place \fu as one of the few actively accreting
sources within the `mass gap' of ${\approx} 2$--5\,$\msun$ where there
are few known massive neutron stars or low mass black holes.  Higher
mass and distance, however, remain viable.
\end{abstract}

\keywords{accretion, accretion disks – stars: low-mass – pulsars:
  general – stars: neutron – X-rays: binaries}

\setcounter{footnote}{0}

\section{Introduction}\label{sec:intro}

Most of the known systems in our Galaxy suspected of harboring a
stellar mass black hole were discovered via observations of X-ray
binary systems that have undergone one or more transient outbursts.
The list of persistently X-ray bright black hole candidate systems is
somewhat shorter, with some of the brightest sources comprised of
systems with high mass ($\gtrsim 3\,\msun$) secondaries, e.g., \cyg
(black hole mass $21.2^{+2.2}_{-2.3}\,\msun$, companion mass
$40.6^{+7.7}_{-7.1}\,\msun$; \citealt{miller-jones:21a}), \lmcthree
(black hole mass $7.0\pm0.6\,\msun$, companion mass
$3.6\pm0.6\,\msun$; \citealt{orosz:14a}), and \lmcone (black hole mass
$10.9\pm1.4\,\msun$, companion mass $31.8\pm3.5\,\msun$;
\citealt{orosz:09a}).  These three systems also spend a significant
fraction of their time in a soft X-ray spectral state (ranging from
$\gtrsim 20\%$ of the time in the case of \cyg,
\citealt{grinberg:13a}, to 100\% in the case of \lmcone).  \fu is a
low mass X-ray binary (LMXB) system, with comparable observed
brightness to \lmcone and \lmcthree, that is also thought to harbor a
black hole \citep{margon:78a,thorstensen:87a,hakala:99a}.  Also
similar to the above high mass X-ray binary (HMXB) sources, it has
frequently been observed in a soft X-ray spectral state, and in fact
has never been identified with a hard X-ray spectral state
\citep[e.g.,][]{yaqoob:93b,ricci:95a,nowak:99c,wijnands:02c,nowak:11b,maitra:14a,maccarone:20a}.
Unfortunately, partly owing to its persistent nature, and thereby not
allowing optical measurements of binary parameters in quiescence,
little is known about the system's mass, distance, and inclination.
Mass estimates in particular have ranged from hypothesizing a neutron
star primary \citep{bayless:10a} all the way to suggesting a black
hole of possibly tens of solar masses \citep{nowak:11b}.

\fu exhibits a 9.33\,hr optical orbital period
\citep{thorstensen:87a}, which would be indicative of a low mass
companion ($\lesssim 1\,\msun$) if Roche Lobe overflow supplies
accretion from a secondary with a radius equal to or greater than that
expected for a main sequence star.  Optical emission is likely
dominated by the accretion disk \citep{hakala:14a}; however, numerous
analyses have attributed observed $\pm20\%$ optical modulation to
irradiation/heating of the surface of the secondary
\citep{bayless:10a,mason:12a,gomez:15a}.  Requiring that irradiation
of the secondary be primarily responsible for the optical modulation
suggests that the secondary subtends a significant fraction of the sky
from the primary's point of view. Thus the binary system would have a
small total mass, perhaps most consistent with a neutron star
primary\footnote{The main arguments against a neutron star primary has
  been the lack of nuclear bursts, pulsar signatures, or the need for
  any `surface emission' component in any analysis of X-ray spectra
  for all the observations that have been performed to date, including
  those presented in this work.} or low mass black hole
\citep{bayless:10a,gomez:15a}.  These optical models also allow for a
range of system inclinations (e.g., \citealt{bayless:10a} fit
inclinations ranging from $20^\circ$--$70^\circ$). On the other hand,
\citet{hakala:14a} model the optical spectral energy distribution as
being due to X-ray heating of the outer accretion disk, and based upon
these models argue for a distance of \emph{at least 15\,kpc} and a
commensurately large mass $>15\,\msun$.  Furthermore, high modulation
amplitudes and complex orbital structure possibly seen in some optical
observations has been attributed to heating of the outer edge of a
disk seen at large (but not eclipsing) inclination angles
\citep{hakala:99a,russell:10a}.

It is known from modeling both cold and hot phase interstellar medium
absorption along the line of sight to \fu that it resides outside of
the plane of the Galaxy at a minimum distance of $\gtrsim 5$\,kpc
\citep{nowak:08a,yao:08a}.  Lying within the Galactic halo means that
any prior radio jet activity from \fu has not interacted with a dense
interstellar medium and therefore has not produced a nearby radio
``hot spot'' that can be confused with current jet emission.  This has
allowed the most stringent upper limits to be placed upon the ratio of
radio to X-ray flux during a black hole ``soft state''
\citep{russell:11a,maccarone:20a}.

\citet{maccarone:20a} also consider a variety of other optical
observations of \fu, including spectroscopic studies of Bowen
fluorescence lines \citep{longa:15a} and parallax and proper motion
studies using the \gaia EDR3 catalog \citep{gaia:16a,gaia:20a}, to
evaluate distance, mass, and potential kick velocity constraints
(assuming a Galactic halo orbit) for this system.  Their most
important constraint is that \gaia results are consistent with a
median distance of 7\,kpc with a 95\% likely range of
3--15\,kpc. Depending upon specific assumptions, these measurements
suggest that \fu may be the fastest moving black hole system known in
the Galaxy, but its mass remains ambiguous. If the Bowen fluorescence
line studies \citep{longa:15a} are taken at face value, then the mass
ratio in the system is $\approx 0.25-0.3$, arguing for a low mass
black hole or high mass neutron star system.

Studies based solely upon X-ray observations have been no more
definitive.  All studies agree that the soft X-ray spectrum is
consistent with emission from an accretion disk
\citep[e.g.,][]{mitsuda:84a} with a high peak temperature and low
normalization \citep[see][]{nowak:11b,maitra:14a}.  Such a high
temperature/low normalization can be achieved by combinations of low
mass, high spin, and high accretion rate for a black hole accreting
from a disk being viewed at high inclination and/or large distance,
but degeneracies in the model fits abound
\citep{nowak:11b,maitra:14a}.  The lack of X-ray eclipses by the Roche
lobe of the secondary limits the disk model inclinations to $\lesssim
80^\circ$ (for a $1\,\msun$ secondary), but otherwise allows a wide
range of parameters.

Taken as a whole, the optical and X-ray results to date allow for, but
do not strictly require, the compact object in \fu to be a low mass
black hole residing in the `mass gap' of $\approx 2$--5\,$\msun$
\citep{farr:11a}, which approximately spans from the masses of the
most massive known neutron stars to the least massive known black
holes.  Few compact objects, especially those that have exhibited
active accretion, have been found within this mass range. It has been
argued that the maximum mass of a neutron star is \emph{at least}
$2.19\,\msun$ ($1\sigma$ confidence, based upon a $2.35\pm0.17\,\msun$
mass estimate for PSR~J0952$-$0607, as well as high mass estimates for
several other well-observed systems; \citealt{romani:22a}).  For low
mass black holes, several non-interacting binaries have been claimed
to harbor black holes with likely masses $<5\,\msun$
($3.3^{+2.8}_{-0.7}\,\msun$ for 2MASS~J05215658+4359220,
\citealt{thompson:19a,thompson:20a}; $4.53\pm0.21\,\msun$ and
$4.4\pm2.8\,\msun$ in the globular cluster NGC~3201,
\citealt{giesers:18a,giesers:19a}; or \citealt{shenar:22a} who
identify ten O-star binaries, out of a sample of fifty one, that might
harbor black holes with masses $<5\,\msun$). Among at least
intermittently active X-ray binaries, XTE~J1650$-$500 could have an
upper mass limit of $\approx 4\,\msun$ if the accretion disk
contributes significantly to the optical lightcurve of the system
\citep{orosz:04a}.  Thus it would be extremely interesting if
observations could determine whether or not \fu is another example of
a low mass black hole system, in this case a persistently accreting
one.

The goal of this work is to consider a new set of X-ray
data\footnote{Of the ten \nustar observations discussed herein,
  however, nine have previously been presented by \citet{sharma:21a}
  and one has been presented by \citep{mudambi:22a}.  None of the
  joint \nicer observations were discussed in either of those works.}
for \fu, and consider them along with \gaia optical constraints
comparable to those discussed by \citet{maccarone:20a}.  We further
include considerations of theoretical disk modeling and observational
X-ray properties for other known black hole systems in order to
reassess constraints on the mass and distance of the \fu system. As we
discuss below, what makes these \nustar observations of \fu unique is
that they extend to energies of 20\,keV or beyond, with high
signal-to-noise and minimal background uncertainties.  This allows
more precise fitting of disk models than we have achieved with
previous studies. Here we find that the \fu spectra are in a regime
with minimal degeneracy as regards the fitted spin and inclination for
the relativistic disk models that we have chosen to consider
\citep[see][]{parker:19a}.

The outline of the paper is as follows.  In \S\ref{sec:context} we
place the historical behavior of \fu in context in relation to other
well-observed black hole systems by presenting a variation of
color-intensity diagrams created from \rossi (\rxte) observations. In
\S\ref{sec:obs} we describe the newer X-ray observations from \nustar
and \nicer that we utilize in this work. \S\ref{sec:timing} and
\S\ref{sec:spectra} describe variability analyses (including
consideration of \maxi lightcurves for \fu) and spectral analyses of
these data, respectively. Included with our discussion of the spectral
fits, we consider in \S\ref{sec:scale} the scaling relationships that
describe degeneracies of the fit parameters.  We then summarize our
conclusions in \S\ref{sec:discuss}.

\section{\fu `\lowercase{q}-diagram'} \label{sec:context} 
\begin{figure}
\begin{center}
\includegraphics[width=0.46\textwidth]{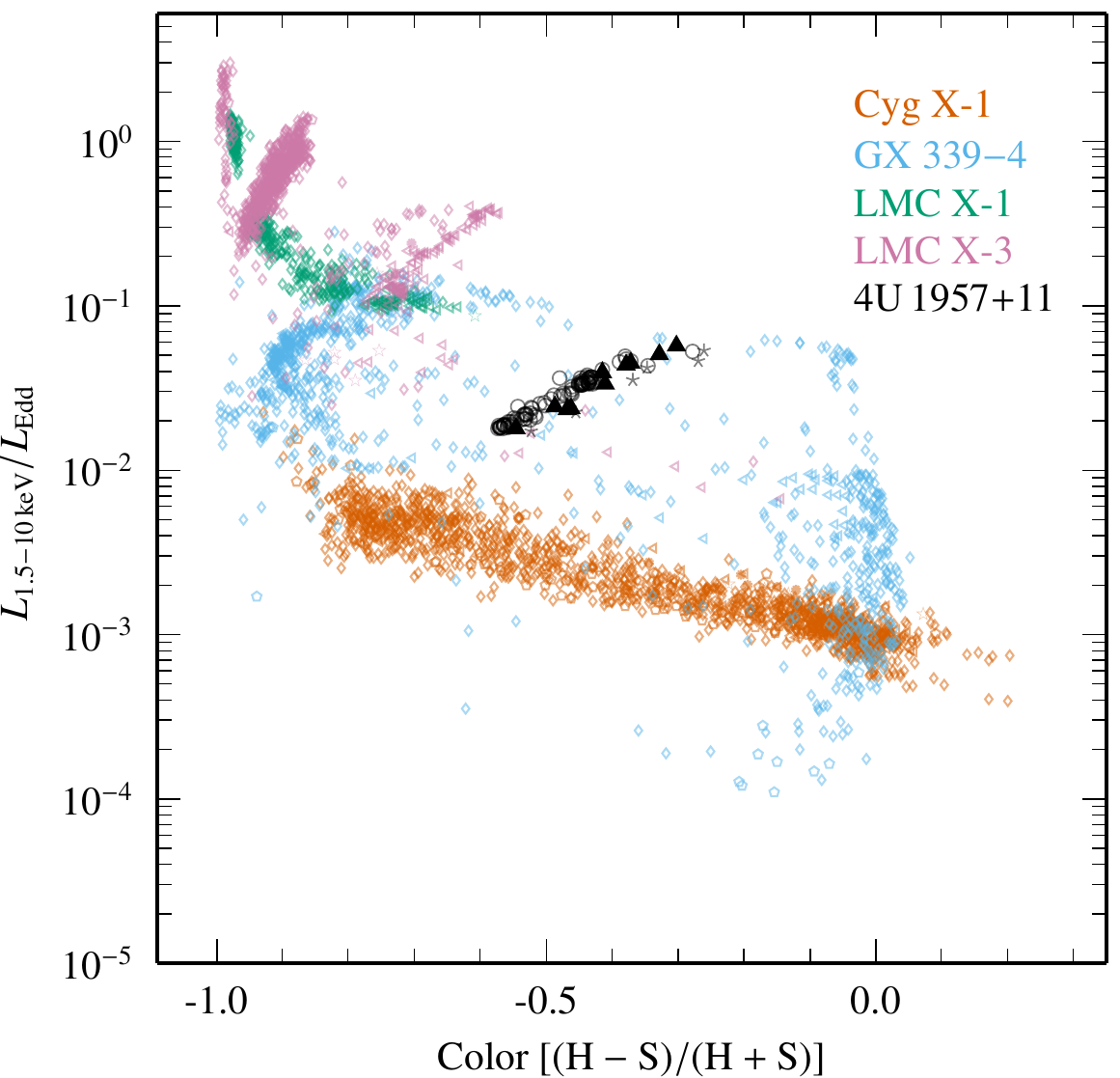}
\end{center}
\caption{A `q-diagram' of intensity vs.\ color for \rxte observations
  of several different black hole candidate systems.  Rather than
  relying on specific spacecraft observed counts, this diagram is
  constructed from model fits to the data, extrapolated to the
  1.5--5\,keV and 5--10\,keV bands and normalized to the Eddington
  ratio for assumed masses and distances (see text).  Pentagons,
  triangles, and diamonds are for \rxte-PCA gain epochs 1, 2, and 5
  (see \citealt{shaposhnikov:12a}, and references therein)
  respectively. Solid grey triangles are for our Comptonized disk fits
  to the \nustar-only spectra, and the solid grey stars are for our
  Comptonized disk fits to the joint \nicer+\nustar spectra (see
  \S\ref{sec:obs}).}
  \label{fig:Q}
\end{figure}

It has been known for nearly thirty years that at low luminosities
black hole systems tend to be in a spectrally hard state.  This
spectrally hard state can persist to high luminosities during the rise
of a transient outburst.  On the other hand, persistent high
luminosity sources (such as \lmcone and \lmcthree) and transients that
reach very high luminosities tend to be in spectrally soft states.
Spectrally soft state transient sources tend to remain in such soft
states until they reach a sufficiently low luminosity where these
systems again switch to spectrally hard states (see, e.g., Figure~1 of
\citealt{nowak:95a} or Figure 2 of \citealt{maccarone:03aa}).  Here
``low'' and ``high'' luminosity should be considered defined in terms
of the black hole's fractional Eddington luminosity.  This behavior
has been encapsulated in what has become known as the ``q-diagram'',
with X-ray color plotted along the horizontal axis (hard relative to
soft counts increasing to the right) and X-ray count rate increasing
along the vertical axis \citep[e.g.,][]{fender:04a}.  Typically this
diagram is plotted for a specific spacecraft, so the definitions of
colors and count rates have not been universal, but characteristic
patterns have emerged.  It has been noted that below $\approx 1\%$ of
a source's Eddington luminosity, black hole systems are spectrally
hard and radio loud, with the transition back to a hard state as the
outburst fades occurring over a much narrower range of luminosities
than the outburst-driven transition from a hard state
\citep{maccarone:03a}.  Winds are seen most often in soft, bright
spectral states and their detection may even be dependent upon system
inclination \citep{ponti:12a}.

To place the behavior of \fu in the context of such a diagram, we have
analyzed the entire \rxte catalog for five black hole sources: the
predominantly soft and persistent sources \lmcone, \lmcthree, and \fu,
and the more spectrally variable sources \cyg and \gx, with the latter
source also being transiently active.  We choose these sources for
several reasons.  First, all have an extensive \rxte database.  All
have shown spectrally soft, disk-dominanted states.  (For the case of
\cyg, based upon the prior work of \citealt{wilms:06a} and
\citealt{grinberg:14a}, we estimate that $\approx 6\%$ of the \rxte
data points are ``soft'' states, and $\approx 4\%$ are
``transitional'' states.) The three HMXB have well-determined
distances and masses, although in at least the case of \lmcone (see
below) their HMXB nature might lead to some peculiarities in the
diagram.  Both \fu and \gx, however, are LMXB, and the latter source
is often used as the ``canonical'' example of the q-diagram.

So as to make this diagram more ``generic'' for observations with
other X-ray missions, we fit a simple spectral model to 3-20\,keV band
Proportional Counter Array (\pca) data: an absorbed disk plus broad
line plus broken powerlaw.  We have previously used this model to
describe phenomenologically \rxte observations of both hard and soft
spectral states of \cyg \citep{wilms:06a}. We form colors by a
comparison of the (absorbed) model energy fluxes in the 1.5--5\,keV
and 5--10\,keV energy bands (i.e., hard minus soft rate, divided by
hard plus soft rate). We choose these bands as each has good overlap
with most modern X-ray/gamma-ray instruments (i.e., the 0.2--12\,keV
of \nicer and the 3--70\,keV of \nustar) and the soft band is not
heavily influenced by absorption for equivalent neutral columns $\nh
\lesssim 10^{22}\,\cmt$.  Thus the hope is that by using these bands
systematic effects will be minimal when comparing results among
different missions.  To further ``normalize'' the q-diagram, we
rescale the fluxes assuming isotropic emission for assumed masses and
distances: \lmcone: 10.9\,$\msun$, 50\,kpc \citep{orosz:09a};
\lmcthree: 7\,$\msun$, 50\,kpc \citep{orosz:14a}; \gx: 5.8\,$\msun$,
6\,kpc \citep{hynes:03a}; \cyg: 21.2\,$\msun$, 2.22\,kpc
\citep{miller-jones:21a}; and \fu: 4.6\,$\msun$, 7.8\,kpc (see
\S\ref{sec:discuss}).  We then ratio this 1.5--10\,keV luminosity to
the Eddington luminosity for the assumed mass.  The results are
presented in Figure~\ref{fig:Q}.

Transitions to a low luminosity, spectrally hard state at luminosities
$\lesssim 1\%\,\ledd$ are seen for \lmcthree, \cyg, and \gx. No such
evidence presents itself for \fu or \lmcone.  For \gx, \lmcthree, and
to a lesser extent \cyg, bright, spectrally soft states are apparent
as upturns on the upper left side of the diagram, with greater extents
and curvature to the right for higher fractional Eddington luminosity.
The two most unusual sources in this regards are \lmcone, which is
spread predominantly horizontally along the diagram, and \fu.  \lmcone
is already known to have slightly unusual X-ray properties for a
``soft state'' black hole system, e.g., peculiar correlations between
normalization and temperature for disk model fits to the spectra, and
high fractional variability (for a disk-dominated soft state) with
little coherence between X-ray variability in the soft and hard X-ray
energy bands \citep{wilms:01a,nowak:01a}.  This has been attributed to
the wind-fed nature of accretion in this system, which may lead to an
extremely small circularization radius for the disk in this system
(see the discussions in
\citealt{wilms:01a,nowak:01a,beloborodov:01a}). \lmcthree and \cyg,
although also HMXB, are likely to have substantially larger disk
circularization radii, and therefore may be more likely to behave in a
similar manner to an LMXB system.

On the other hand, the shape and extent of the track exhibited by \fu
is not markedly different than the multiple soft state tracks seen for
\lmcthree, or those seen for \gx (and seen weakly for \cyg), except
for the fact that it is noticeably harder.  \lmcthree, \gx, and \cyg
only achieve such hardnesses while in or in transition to a low
luminosity hard state.  As we shall argue below, this unusual spectral
hardness can be the result of two potential properties of \fu: high
spin and high inclination.  For our hypothesized spectral model when
considering \nustar spectra, these fit parameters are not expected to
be subject to strong fitting degeneracies \citep{parker:19a}.

\begin{deluxetable*}{rcccccccccccr}
\setlength{\tabcolsep}{0.05in} \tabletypesize{\footnotesize}
\tablewidth{0pt} \tablecaption{Observation Log}
\tablehead{\colhead{Epoch} & \colhead{Date} & \colhead{\nustar}
  & \colhead{Start} & \colhead{Stop} & \colhead{GTI} &
  \colhead{\nicer} & \colhead{Under/Over} &
  \colhead{Start} & \colhead{Stop} & \colhead{GTI} 
  & \colhead{3-70\,keV Flux} & \colhead{Color} \\ & &
  \colhead{ObsID} & \colhead{(days)} & \colhead{(days)} &
  \colhead{(ks)} & \colhead{ObsID} & & \colhead{(days)} &
  \colhead{(days)} & \colhead{(ks)} 
  & \colhead{($\mathrm{erg/cm^{2}/s}$)}}
  \startdata 
1 & 16-11-2013 & 30001015002 & 6612.65 & 6613.94 & 50.0 
   & \nodata & \nodata/\nodata & \nodata & \nodata & \nodata 
   & $0.51\times10^{-9}$ & Pink \\ 
2 & 16-09-2018 & 30402011002 & 8377.31 & 8378.14 & 37.2 
   & 1100400103 & 200/1.0 & 8378.09 & 8378.17 & 1.7
   & $0.35\times10^{-9}$ & Brown \\ 
3 & 13-03-2019 & 30402011004 & 8555.42 & 8556.26 & 33.5 
   & \nodata & \nodata/\nodata & \nodata & \nodata & \nodata 
   & $0.87\times10^{-9}$ & Gold \\ 
4 & 29-04-2019 & 30502007002 & 8602.87 & 8603.37 & 20.7 
   & 2542010101 & 200/1.0 & 8602.61 & 8602.94 & 4.5 
   & $0.55\times10^{-9}$ & Orange \\ 
   \nodata & \nodata & \nodata & \nodata & \nodata & \nodata 
   & 2542010102 & 200/1.0 & 8603.00 & 8603.39 & 3.8 
   & \nodata & \nodata \\ 
5 & 15-05-2019 & 30402011006 & 8618.50 & 8619.54 & 37.0 
   & \nodata & \nodata/\nodata & \nodata & \nodata & \nodata 
   & $1.15\times10^{-9}$ & Blue \\ 
6 & 04-06-2019 & 30502007004 & 8638.84 & 8639.28 & 20.1 
   & 2542010201 & 200/1.0 & 8638.88 & 8638.98 & 3.8 
   & $1.10\times10^{-9}$ & Green \\ 
   \nodata & \nodata & \nodata & \nodata & \nodata & \nodata 
   & 2542010202 & 200/1.0 & 8639.02 & 8639.29 & 5.8 
   & \nodata & \nodata \\ 
7 & 19-07-2019 & 30502007006 & 8683.26 & 8683.74 & 18.4 
   & 2542010301 & 200/1.0 & 8683.40 & 8683.93 & 14.2 
   & $1.53\times10^{-9}$ & Cyan \\ 
8 & 10-09-2019 & 30502007008 & 8736.08 & 8736.45 & 10.1 
   & 2542010401 & 200/1.0 & 8736.09 & 8736.50 & 10.0 
   & $1.76\times10^{-9}$ & Purple \\ 
9 & 20-10-2019 & 30502007010 & 8776.54 & 8776.98 & 19.6 
   & 2542010501 & 200/1.0 & 8776.54 & 8776.94 & 11.7
   & $0.89\times10^{-9}$ & Black \\ 
10 & 30-11-2019 & 30502007012 & 8817.88 & 8818.32 & 20.5 
   & 2542010601 & 300/1.5 & 8817.84 & 8817.98 & 4.0 
   & $0.51\times10^{-9}$ & Magenta \\ 
   \nodata & \nodata & \nodata & \nodata & \nodata & \nodata 
   & 2542010602 & 300/1.5 & 8818.03 & 8818.30 & 6.2 
   & \nodata & \nodata  \\ 
\enddata 
\tablecomments{Times refer to spacecraft times, and have not been
  barycenter corrected.  Good Time Intervals (GTI) first start and
  final stop times are relative to Modified Julian Date (MJD) 50,000,
  while the summed GTI exposure is given in ks. Under/Over refers to
  the filter criteria applied to the \nicer data, where the acceptable
  \texttt{underonly\_range} was set between 0 and the given value and
  the acceptable \texttt{overonly\_range} was set between 0 and the
  given value.  (The related \texttt{overonly\_expr} was also scaled
  based upon this latter upper value. Color refers to colors used (in
  online version) in subsequent plots.)}\label{tab:obs_log}
\end{deluxetable*}

\section{Observations}\label{sec:obs}

In this section, we describe the observations of \fu performed with
several different instruments. The main focus of this paper will be on
combining information from the \textsl{Neutron Star Interior
  Composition Explorer} (\nicer; \S\ref{sec:nicer};
\citealt{gendreau:16a}) and the \textsl{Nuclear Spectroscopic
  Telescope Array} (\nustar; \S\ref{sec:nustar};
\citealt{harrison:13a}).
  
As of 2022 November, there have been 10 observations of \fu taken with
the \nustar observatory. The earliest of these observations was
conducted quasi-simultaneously with both \textsl{XMM-Newton} and
\textsl{Hubble Space Telescope-Cosmic Origins Spectrograph}
observations.  As that is a unique combination of observatories, with
their own systematics to consider not represented in our other \nustar
observations, we defer discussion of the non-\nustar spectra to a
future paper.  (Analyses of those data, however, do not fundamentally
alter any of the conclusions discussed in this work.)

Seven of the \nustar observations were conducted quasi-simultaneously
with \nicer observations. All of the \nustar observations showed \fu
in a seemingly disk-dominated, spectrally soft state, as has been
typical for this source.  In what follows we first consider all 10
\nustar observations analyzed by themselves, and then the subset of
seven \nustar observations conducted quasi-simultaneously with \nicer
observations analyzed jointly as a group. We also utilize observations
from the Monitor of All-sky X-ray Image (\maxi;
\citealt{matsuoka:09a}) to place our pointed observations in the
context of the long term behavior of the source.

\subsection{\nustar Observations}\label{sec:nustar}
The \nustar observatory nominally covers a range of 3--79\,keV with
its two focal plane module instruments (\fpma and \fpmb) designed to
record the energy, arrival time, and location of each incoming
event. There have been ten epochs of observations of \fu with \nustar;
observation times and integrated exposures are listed in
Table~\ref{tab:obs_log}.

To process our \nustar observations of \fu, we utilized the
\textit{nupipeline} and \textit{nuproducts} tools from \heasoft, with
standard parameter choices and filter criteria.  We used calibration
products current as of 2021 November (release
\texttt{20211020}). Spectra and lightcurves were extracted from
circular regions with diameters of 50\,arcsec centered on the image of
the source, while backgrounds were extracted from source free,
circular regions on the same chip with diameters of 100\,arcsec.  The
lightcurve bin time was chosen to be 1\,msec.

For spectral analyses, the pipelines created both \rmf and \arf
standard response matrices.  In all spectral analyses, we kept the
\fpma and \fpmb spectra separate, but grouped them to a common energy
bin grid.  Specifically, we grouped the data during spectral analysis
with \textsl{Interactive Spectral Interpretation System} (\isis)
v1.6.47 using the \texttt{group} function simultaneously applied to
the \fpma and \fpmb spectra.  We chose a minimum combined
signal-to-noise of $\sqrt{2}*5$ per grouped bin (i.e., an average of
25\,counts per bin per detector, in spectral regions with low
background), and a minimum of 2--10 detector channels (increasing by
1~channel at each step) per bin at energies starting at (3, 7, 12, 19,
27, 37, 48, 61, 75)\,keV.  The latter grouping criteria are designed
to slightly oversample, but roughly follow, the detector resolution as
a function of energy. The ``resolution criteria'' in fact dominates
the binning at all but the highest energies.  We then noticed bins
that fully fell within the 3--100\,keV range, but the final energy bin
always encompassed a wide energy range with an upper value
$>100$\,keV, due to the signal-to-noise criteria, and was ignored. The
upper limit of our fits therefore range from $\approx 20$--70\,keV.

For all figures in this work we combine the \fpma and \fpmb spectra
for presentation purposes, and these spectra are shown ``flux
corrected'' using only the \nustar response matrices without reference
to any specific model.  (Specifically, we use the \isis
\texttt{plot\_unfold} function, available via the \texttt{isisscripts}
package at Remeis Observatory; see the description in
\citealt{nowak:05a}.)

\subsection{\nicer Observations}\label{sec:nicer}

The \nicer instrument is composed of 56 separate optics and detectors,
52 of which are operational and are used in our spectral and timing
studies.  It allows for very precise timing analysis with precision of
better than 300 nanoseconds over the range of 0.2--12\,keV. There have
been fourteen individual (non-zero exposure time) epochs of
observations of \fu taken with the \nicer instrument. Seven of these
epochs (several of which consist of more than one observation ID) were
performed quasi-simultaneously with \nustar.  Dates and total
integrated observing times are presented in Table~\ref{tab:obs_log}.

In order to create ``level 2'' data products, the \nicer data were
processed with the \texttt{nicerl2} tool from \heasoft and the
calibration database current as of 2021 November (release
\texttt{xti20210707}). Default filter criteria were employed, with the
exception of filtering on ``undershoot'' and ``overshoot'' events (see
\citealt{gendreau:16a}).  Prior to the release of \heasoft, the
standard filter criteria for these values led to exclusion of a large
fraction of events in a few of our observations.  A modest loosening
of the undershoot/overshoot limits, which has now become standard in
\nicer analyses, recovered a substantial number of events. The
specific values that we used to filter the data are also listed in
Table~\ref{tab:obs_log}.

\begin{figure*}
\begin{center}
\includegraphics[width=0.99\textwidth,viewport=32 0 419 96]{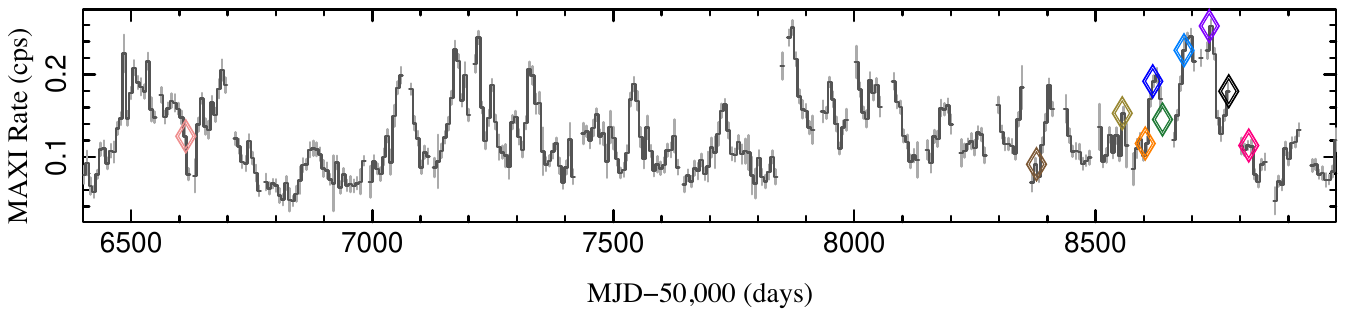}
\end{center}
\caption{\maxi lightcurve for \fu covering a period spanning the
  \nustar/\nicer observations, using 7\,d bins and weighted means and
  $1\sigma$ error bars.  Bins are included only if they are comprised
  of ten or more individual observations.  Times, and overlapping
  weighted mean \maxi flux (linearly extrapolated if in a data gap),
  of the pointed \nustar observations are indicated by the hollow
  diamonds, color-coded (online version) to match the \nustar data in
  subsequent plots.} \label{fig:maxi}
\end{figure*}

Additionally, we use the customized per observation (as opposed to
``canned'') response matrices that became available with the \heasoft
release, using the \nicerarf and \nicerrmf tools and the above cleaned
event files entered as inputs.  Backgrounds for the \nicer
observations were estimated using \texttt{v5} of the
\emph{nibackgen3C50} tool. Background, however, did not play a large
role in our \nicer spectral analyses.

For spectral analyses, we grouped the \nicer spectra based upon
spectral resolution, using 1, 2, 3, 4, 5, and 6 channels per bin
starting at 0.4, 1.0, 2.4, 4.0, 6.4, and 9.0\,keV, respectively.  We
included energy channels with boundaries fully within the 0.4--10\,keV
range.  As for the \nustar spectra, whenever there are multiple
spectra during a given epoch, we combine the \nicer spectra for
presentation purposes, but kept them separate during analysis.

\section{Timing Analysis} \label{sec:timing}

\subsection{Long Term Analysis -- \maxi Observations}\label{sec:maxi}

For long-term behavioral analysis of \fu, we utilized the
\textsl{Monitor of All Sky X-ray Image} (\maxi) 1.5\,hr cadence
observations which cover an energy range of 2--20 keV.  \maxi is
hosted on the International Space Station, and it is composed of two
semi-circular X-ray cameras, a Gas Slit Camera and the Solid-state
Slit Camera, and is designed to observe the sky with wide fields of
view, although data gaps can occur due to blockage by the space
station, the earth, instrumental issues, etc. \citep{matsuoka:09a}. We
present the 2--20\,keV \maxi lightcurve in Figure~\ref{fig:maxi}
for the period covering our \nicer and \nustar observations. We have
binned the data into 7\,d bins using an error weighted mean.  We only
include data bins that have at least ten contributing measurements,
although we do not impose any restriction on the minimum time span
covered by the measurements.  We also indicate the time and
overlapping weighted mean \maxi flux for the pointed \nustar
observations.  (We use a linearly extrapolated value of the \maxi flux
for those \nustar observations that fall within a data gap.)

The lightcurves for the full \maxi lifetime (through 2022 November;
not shown) span approximately a factor of seven from the lowest to
highest rates, with the extrapolated \maxi rates at the times of our
\nustar observations spanning a factor of 2.8.  Our brightest
observations overlap with the highest rates observed by \maxi, and
there is perhaps a further factor of 2.5 range at lower \maxi rates.
Our pointed observations therefore span slightly more than half the
dynamic range exhibited in the \maxi lightcurve, with approximately
23\% of the \maxi bins having lower rates than our faintest
observation.  We thus have a reasonably broad and diverse sample of
the historical spectral behavior exhibited by \fu.  There are no
indications within the \maxi lightcurve (nor within the \rxte
generated q-diagram shown in Figure~\ref{fig:Q}) for a transition to a
spectrally hard state at low observed rates.  Given the extremely high
fitted disk temperatures discussed below, such a transition could be
difficult to discern solely from these color-intensity diagrams.

\begin{figure}
\begin{center}
\includegraphics[width=0.45\textwidth,viewport=65 10 580 518]{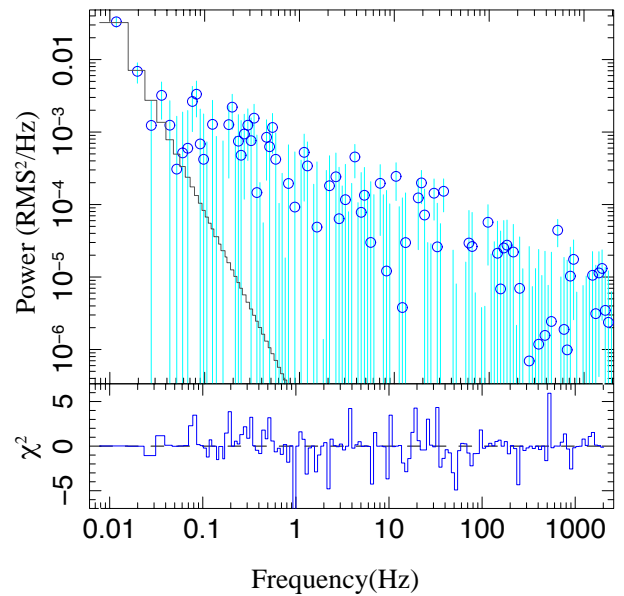}
\end{center}
\caption{Power spectral density (PSD) for the 2--8\,keV lightcurve of
  \nicer ObsID 2542010401.  The PSD here is fit with a constant, to
  describe the Poisson noise level, and a powerlaw.  The constant
  component has been subtracted, leaving an ``upper envelope''
  following the expected $P(f) \propto f^{-0.5}$ residuals from
  Poisson noise fluctuations as high frequency residuals. The
  intrinsic variability of the source is $1.9\%\pm0.2\%$ root mean
  square (rms).}
\label{fig:psd}
\end{figure}

\subsection{Short Term Analysis} \label{sec:short}

We measure the fractional variability of the individual observation
lightcurves via their Power Spectral Densities (PSD), calculated via
Fast Fourier Transforms (FFTs). We normalize these PSD following
\citet{belloni:90a} such that integrating over Fourier frequency
yields the squared fractional variation of the counts per unit
frequency.

The PSD for each \nicer and \nustar observation was composed of an
average of each of the individual PSD for the good time intervals
(GTI) of each epoch of observations. The PSD were logarithmically
averaged over frequencies with bins having $\Delta f/f \approx0.08$.
The most significant variability was found for the \nicer lightcurves
(and \nicer timing analysis is more straightforward than \nustar
timing analysis; see \citealt{bachetti:15a}), so here we discuss those
results. The high timing precision of the \nicer instrument allowed us
to investigate each observation over the range of
$2^{-7}$\,Hz--2048\,Hz. We explored PSD in various energy bands, but
the fractional variability was always small and the intrinsic PSD is
above the modeled Poisson noise level only in the $\approx
0.01$\,Hz--0.1\,Hz range.  We found the highest fractional variability
in the 2--8\,keV band, so we discuss those results here.  (Further
restricting the energy band to even higher energies results in worse
signal-to-noise, owing to a decrease in the \nicer effective area at
those energies.) We used \isis to fit a \powerlaw plus \constant
function (the latter to represent the PSD of the Poisson noise), with
the \powerlaw then being integrated over the above frequency range to
measure the root mean square (rms) variability.

Overall, we found that \fu was remarkably quiet in all observations,
with rms variabilities ranging from undetected (90\% confidence upper
limits of 0.7\%) to $2\%$.  The two most significant detections are
for \nicer ObsID 2542010201 (rms $2.0\%\pm0.5\%$, 90\% CL) and
2542010401 (rms $1.9\%\pm0.2\%$, 90\% CL).  ObsID 2542010401 is among
the brightest states that we observed with one of the most significant
hard tails; however, ObsID 2542010201 is slightly fainter with a
weaker hard tail.  In all cases, however, the observed variability is
modest, as shown in Figure~\ref{fig:psd}. For comparison, in its soft
state \cyg can reach rms variability of up to 20\%
\citep{pottschmidt:00a,axelsson:05a,axelsson:06a,grinberg:14a}.  In
its hard state, \cyg achieves even higher rms variability of almost
40\%, and has sufficient signal-to-noise so as to observe multiple
Lorentzian structures in the PSD \citep[see,
  e.g.,][]{nowak:00a,pottschmidt:02a,axelsson:05a,grinberg:14a}.  Even
with the excellent effective area of \nicer, we are unable to
characterize the PSD beyond a simple, weak powerlaw as has been seen
historically for soft state black hole systems (e.g.,
\citealt{miyamoto:94a}).

\section{Spectral Fits} \label{sec:spectra}

Based upon prior work we consider fits that model the spectra as being
dominated by a disk with the addition of a modest hard spectral tail.
We present fits for the \nustar-only spectra (ten epochs) and the
joint \nicer/\nustar spectra (seven epochs).  For the former we
utilize a Comptonization model with `disk' seed photons (\eqpair;
\citealt{coppi:92a}), similar to the analysis of \fu presented by
\citet{nowak:11b}, and a relativistic disk model (\polykerrbb;
\citealt{parker:19a}, based upon \kerrbb; \citealt{li:05a}) modified
by Comptonization (\thcomp; \citealt{zdziarski:20a}).  For the joint
\nicer/\nustar spectra we present only the \thcomp$\otimes$\polykerrbb
models.  We begin, however, with a discussion of the scaling
relationships we expect for the dominant disk component of the
spectrum.  This is an extension of previous discussions of such
scaling relationships presented by \citet{nowak:11b}. We modify this
discussion, however, based upon the fact that the \nustar spectra
strongly constrain both the fitted black hole spin and disk
inclination for our chosen disk model, as was suggested would be the
case for such high signal-to-noise and broad-band spectra by
\citet{parker:19a}.

\begin{figure}
\begin{center}
\includegraphics[width=0.45\textwidth,viewport=45 32 554 404]{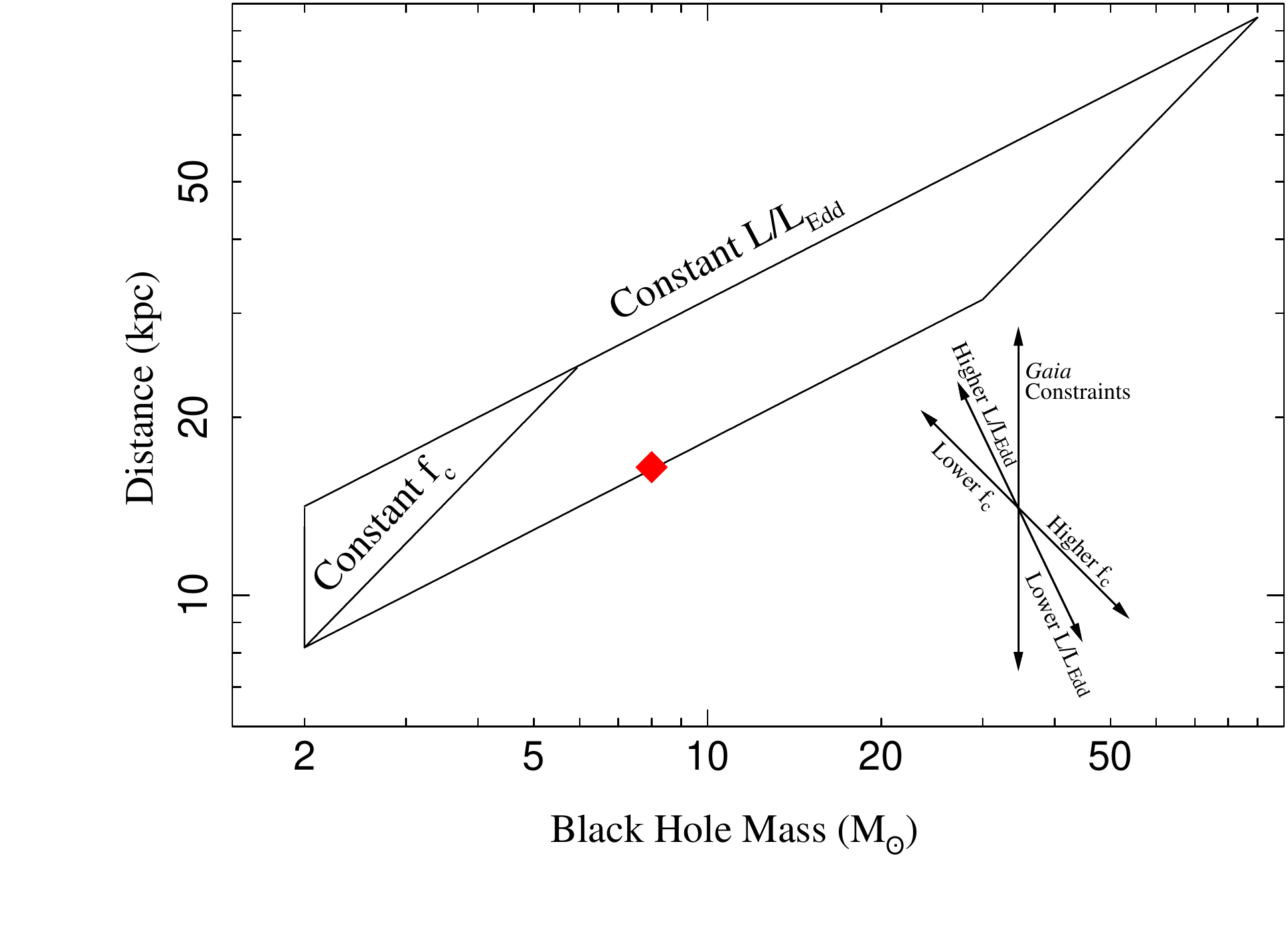}
\end{center}
\caption{Lines highlighting portions of the degenerate parameter space
  for our spectral fits.  Moves along lines of $D \propto M^{1/2}$
  correspond to degenerate fits with constant $\lerat$ but $\fc
  \propto D^{1/2} \propto M^{1/4}$.  The lower of these lines
  corresponds to $\lerat \approx 0.03$ for the faintest observation,
  while the upper line corresponds to $\lerat \approx 0.1$ for the
  faintest observation.  (The brightest observation is approximately
  five times brighter than the faintest observation.)  For the lines
  of $M \propto D$, degenerate fits have constant $\fc$ and $\lerat
  \propto M \propto D$. Moves along lines of constant $M$ have $\lerat
  \propto D^2$ and $\fc \propto D^{-1/2}$.  The left vertical line at
  $M=2\,\msun$ corresponds to the lowest mass one might consider for a
  stellar mass black hole.  The red diamond corresponds to the mass
  and distance that we have used in our spectral fits employing disk
  models.} \label{fig:constrain}
\end{figure}

\subsection{Scaling Relationships} \label{sec:scale}

Disk models --- with the inclusion of components to describe a hard
X-ray powerlaw tail --- provide excellent descriptions of our data.
The primary parameters of these models relate to the location of the
peak of the spectrum and its overall normalization.  In the models
that we employ, the former is typically characterized by a color
temperature, $\tc$, and the latter is related to the source flux, $F$.
Color-temperature and flux are essentially `fixed' by the
observations; however, their relation to physical parameters of
interest, specifically compact object mass, $M$, compact object spin,
$a^*$, source distance, $D$, accretion disk inclination to our line of
sight, $i$, and mass accretion rate, $\dot M$, exhibit a number of
degeneracies.

In the absence of relativistic effects and when disk emission
dominates the spectrum, one expects that
\begin{equation}
    F \propto \left ( \frac{\rd}{D} \right )^2 \teff^4 \cos i 
      \propto \left ( \frac{M}{D} \right )^2 
              \left ( \frac{\tc}{\fc} \right )^4 \cos i ~~,
\end{equation}
where $\rd$ is a characteristic disk radius $\propto M$, and $\teff$
is the disk peak effective temperature which is related to the color
temperature by a color-correction factor, $\fc$, with $\tc \equiv \fc
\teff$.

As discussed by \citet{parker:19a}, and as we find below for our
\nustar observations, broad-band X-ray spectral fits potentially can
strongly constrain disk inclination and spin.  This is especially true
for regions of parameter space that correspond to high spin and high
inclination since the \emph{width} of the temperature distribution in
the disk (due to relativistic beaming and gravitational redshift) is
\emph{broadest} at high spin and inclination.  The fitted \emph{peak}
of the disk temperature distribution is still highly degenerate as
regards mass, distance, accretion rate, and color-correction factor,
but its breadth helps fix the spin and inclination.

We expect our spectral fits to yield $\fc \propto (M/D)^{1/2}$
$\cos^{1/4} i$.  We will show below that the fitted dependence of
$\fc$ upon $\cos i$ is predominantly systematic.  That is, for a
fixed, assumed $(M/D)$ our fit to $\cos i$ depends upon which data we
include, which in turn constrains the fitted value of $\fc$. The
dependence of $\fc$ upon mass and distance, however, is truly
degenerate in that rescaling $\fc \propto (M/D)^{1/2}$ yields exactly
the same spectrum.  Our spectral fits thus constrain the ratio of
$(M/D)$ only to the extent that we have independent estimates of
$\fc$, e.g., from a theoretical prior obtained from disk atmosphere
models \citep{davis:05a,davis:06a,davis:06b},

We can incorporate other priors to further constrain degenerate
parameters.  In addition to $\tc$, $a^*$, and $\cos i$ being fixed by
fits to the observation, so is the flux, $F$.  This latter quantity
can be related to the fractional Eddington luminosity as $\lerat
\propto F D^2/M$.  Combining with the mass/distance dependence of the
color correction factor, we have
\begin{equation}
     M \propto \fc^{4} \frac{L}{\ledd} ~~, ~~ 
     D \propto \fc^{2} \frac{L}{\ledd} ~~.
\end{equation}
Thus priors on color correction factor and fractional Eddington
luminosity can be translated into constraints on mass and distance.

If we assume a fixed fraction of the Eddington luminosity, $\lerat$,
for a given observation (e.g., hypothesize that the faintest
observation, not showing a spectral transition to a spectral hard
state, is $\sim 0.01 \ledd$) we then have
\begin{equation}
    \tc = \fc \teff \propto \left ( \frac{\dot M}{M^2} \right )^{1/4} 
                    \propto \fc {\dot M}^{1/4} ~~.
\end{equation}
The last proportionality is given by assumption that $(\lerat) \propto
({\dot M}/M)$ is fixed.  Under these assumptions, since $L \propto
D^2$, then $D\propto \sqrt{M} \propto \sqrt{\dot M}$.  Spectral fits
will be degenerate at constant $\lerat$ along lines where $M \propto
D^2$ and $\fc \propto D^{1/2} \propto M^{1/4}$.

If we instead require that the color-correction factor $\fc$ is fixed,
then the fits will be degenerate along lines $M\propto D$, with
$\lerat \propto M \propto D$.  If we move along lines of constant $M$,
then degenerate fits will scale as $\lerat \propto D^2$ and $\fc
\propto D^{-1/2}$. We illustrate these lines of degeneracy in
Figure~\ref{fig:constrain}.

We thus only need to consider spectral fits for a single fiducial mass
and distance, which leads to an assumed set of fractional Eddington
luminosities and fitted probability distributions for the
color-correction factor.  We also obtain fitted distributions for spin
and inclination, but those are independent of our assumed mass and
distance, whereas the overall scale of the color-correction factor
depends upon our assumptions as outlined above.  To the extent that we
can place priors on color-correction factor (via theoretical
expectations), fractional Eddington luminosity (via comparison to
other known black hole system behavior), and source distance (e.g.,
from \gaia measurements), we can place fit constraints on the system
mass and distance.

\subsection{\nustar-only Spectral Fits} \label{sec:nustaronly}

\begin{figure}
\begin{center}
\includegraphics[width=0.46\textwidth,viewport=85 20 580 385,clip]{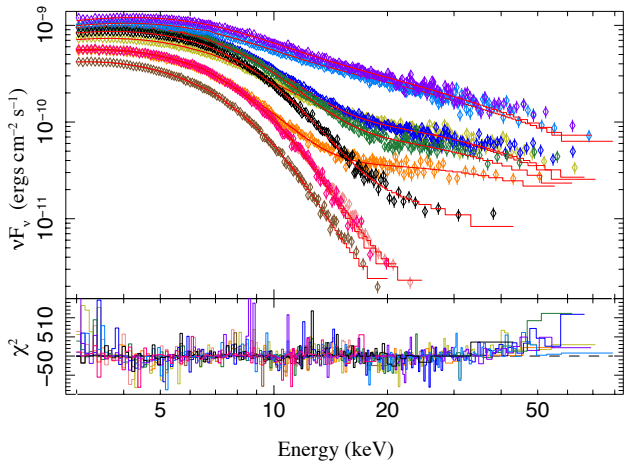}
\end{center}
\caption{Flux-corrected spectra and fits, using an absorbed \eqpair
  model plus relativistic reflection (see text), for \nustar
  observations of \fu.  The individual fits are independent of one
  another.  Color-coding shown in the online color version of this
  figure for the individual spectra (and listed in
  Table~\ref{tab:obs_log}) is used consistently in subsequent figures,
  e.g., showing parameter correlations.}
\label{fig:nustar_spectra}
\end{figure}

In order to be able to directly compare results to \citet{nowak:11b},
we first fit the individual \nustar spectra with a model that consists
of interstellar absorption modifying the \eqpair Comptonization model
plus a gaussian line, specifically: \texttt{tbabs*(eqpair+gaussian)},
The \tbabs model is the current version from the work of
\citet{wilms:00a}, and we use the abundances from that work and the
cross sections of Verner. Given that the equivalent neutral column
along our sight line to \fu is small and the lower cutoff of the
\nustar spectra is 3\,keV, we fix the value to
$1.7\times10^{21}\,\cmt$ (i.e., as found by \citealt{nowak:11b}).  The
gaussian line is constrained to have an energy that lies within
6.2--6.9\,keV and have a width $\sigma <0.3$\,keV, so as not to become
broad and strong enough to falsely fit the continuum spectrum.  For
the \eqpair model we choose seed photons from a disk (essentially the
\diskpn model of \citealt{gierlinski:99a}), fix the seed photon
compactness parameter to one, and then fit the peak disk temperature,
$\ktdisk$, coronal compactness parameter, $\lc$, and coronal seed
optical depth, $\tau_\mathrm{seed}$.  The \eqpair model also provides
a simple reflection model (\emph{without} an Fe line), which we fit
with a `reflection fraction', $R$, constrained to be between 0--2.
(This parameter is typically either weak and/or poorly constrained in
our fits.) To account for differences between the two \nustar focal
plane detectors, we introduce a cross-normalization constant by
multiplying the model by $(1 \pm c_\mathrm{A})$, using the $+$ and $-$
for the \textsl{NuSTAR}-\fpma and \fpmb spectra, respectively.  (That
is, we normalize the fits to the average of these two spectra.)
Results for these fits are presented in Figure~\ref{fig:nustar_spectra} 
and Table~\ref{tab:eqp}.

For these spectral fits, and all spectral fits presented in this work,
parameter tables present values for the fit minimum $\chi^2$ fits.
Error bars, however, are derived from Markov Chain Monte Carlo (MCMC)
analyses using an \isis implementation of the \emcee algorithms of
\citet{foreman:13a} and \citet{goodman:10a}.  Specifically, we evolve
a group of ``walkers'' --- in our case 5 (for the Comptonized
relativistic disk fits discussed below) or 10 walkers (for these
individual \eqpair fits) per free parameter --- for 20\,000 steps, and
create probability distributions for the parameters from the last
10\,000 steps.  We create one dimensional parameter probability
distributions by marginalizing over the other parameters, and then
take the parameter error limits as the bounds that encompass the 90\%
of the probability distribution surrounding the median parameter
value.  (This median value could be different from the ``best fit''
value, but is often nearly identical.)  In some cases the probability
distributions include the fixed bounds that we have imposed (e.g., as
for the gaussian line described above), and those become part of the
90\% probability bounds quoted in the tables.

% Put Table Comments for Below Table Here: 
\newcommand\tcomeqp{For all fits, the equivalent neutral column was
  frozen to $\nh = 1.7\times10^{21}\,\cmtwo$. $c_\mathrm{A}$ is a
  cross normalization constant such that we multiply the model by
  $(1\pm c_\mathrm{A})$, using the $+$ and $-$ for the \nustar-\fpma
  and \fpmb spectra respectively.  Error bars are 90\% confidence
  level.}  \input{table_nustar_eqpair}

These MCMC analyses have implicitly employed non-uniform priors for
some of the parameters.  Uniform priors only apply to those parameters
directly used in the fits.  In Table~\ref{tab:eqp}, the Compton-$y$
parameter is derived from the equilibrium coronal temperature and
final electron/positron optical depth, both of which are themselves
dependent upon the other fitted parameters such as disk temperature,
coronal compactness, and seed optical depth. In principle one can
incorporate probability priors on the independent variables so as to
yield a uniform prior on the dependent variable.  For the case of the
Compton-$y$ parameter this would be a complex function that would be
difficult to incorporate, even numerically.  Below, where we discuss
how we parameterize and fit black hole spin, the required prior is
very slowly changing over the range of spins derived from the
parameter posterior probability distributions, and therefore is only
expected to have small impact on the final error bars.

These individual fits overall describe the \nustar spectra well.  The
fitted peak disk temperatures range from 1.28\,keV--1.52\,keV, which
are rather high values for a soft state black hole and comparable to
the results presented by \citet{nowak:11b}.  Normalizations are also
comparable to the values found by \citet{nowak:11b}, and as discussed
in that work would imply some combination of low black hole mass,
large distance, high inclination, and/or high black hole spin.  We
find, however, more variation in the value of this normalization than
for prior studies of \fu.  Finally, the coronal component in most
cases is fairly weak, as evidenced by small values of the coronal
compactness parameter, $\lc$, as well as small values of the
Compton-$y$ parameter. The latter is essentially the average change in
photon energy due to Comptonization, and peaks at $\approx 5\%$ for
cases where $\approx 40$\% of the photons (based upon net coronal
optical depth) undergo scattering.  We discuss these results in more
detail in \S\ref{sec:discuss}.

We next turn to the Comptonized relativistic disk fits of the
\nustar-only spectra.  In this case, rather than fitting the spectra
individually with \emph{independent} models, we simultaneously fit all
spectra with a uniform black hole mass, spin, distance, and disk
inclination.  Given the fit degeneracies as discussed above, we fix
the black hole mass and distance to 8\,$\msun$ and
16.47\,kpc\footnote{Although this specific distance might at first
  glance seem unusual, it was chosen because for fits using earlier
  versions of the \nustar response matrices and a subset of the
  spectra, this mass and distance yielded a fitted color-correction
  factor of 1.7 for the faintest \nustar observation.  We in fact fit
  our spectra with scaling ``dummy parameters'' that allow us to
  easily replicate the degeneracies in the fits for different masses,
  distances, accretion rates, and color-correction factors, using the
  scaling relations discussed in \S\ref{sec:scale}.}, but allow spin
and inclination to be variable fit parameters.  Rather than fit the
spin directly, we fit a dummy parameter which we tie to the spin via
the relation $a^* = \mathrm{tanh}(p-\mathrm{atanh}(0.998))$, such that
quasi-linear changes in $p$ lead to `interesting' changes in $a^*$
(e.g., $a^*=0.998$ is very different spectrally from $a^*=0.9$, but
$a^*=0$ is not very different spectrally from $a^*=0.1$, and we wish
the fit parameter to reflect this fact across the whole range of
retrograde to prograde spins; see \citealt{maitra:14a}).  For the disk
models, we further use the disk atmosphere color-correction factor as
a fit parameter via direct fitting of this parameter for Epoch 2, and
`dummy parameters' that scale the remaining color-correction factors
to this value. The only other disk parameter that varies among the
individual observations is the disk accretion rate, also fit via dummy
parameters that scale the accretion rates to that from Epoch 2.

For the hard tail we convolve the relativistic disk model with the
\thcomp Comptonization model \citep{zdziarski:20a}. We fix the coronal
temperature to 100\,keV. The remaining fit parameters for each
individual observation are the hard tail photon indices,
$\Gamma_{tc}$, and the corona covering factors, $f_{tc}$. Since we are
fixing the coronal temperature (never having coverage $\gtrsim
70$\,keV, and minimal spectral curvature in the hard tails at lower
energies, this parameter would be very poorly constrained), $f_{tc}$
should not be viewed as purely a covering fraction, but rather as a
proxy for a combination of coronal covering fraction and optical
depth.

The \eqpair fits discussed above indicate that the iron line is weak,
but plausibly present (similar to the possible detections of such a
feature in the \suzaku spectra; \citealt{nowak:11b}).  We model
possible reflection and a broad iron line feature using the \relxillCp
model \citep{garcia:14a}. The \texttt{relxillCp} spin parameter is
tied to the globally fit spin parameter.  We tie the \relxillCp
illuminating photon index to the value of $\Gamma_{tc}$ for each
observation, and consider only the reflected part of the spectrum
(i.e., the directly viewed hard tail only comes from the \thcomp
component).  We assume that the hard tail illuminates the entire disk,
and fit a single, common emissivity for all observations, which
typically pegged at our imposed limit of $\epsilon_\mathrm{rx}=5$).
We use an Fe abundance of $\mathrm{A_{Fe}}=1$.  The remaining fit
parameters are the \relxillCp normalization and ionization parameters,
both of which we fit for each individual observation.  As for the
\eqpair fits, we incorporate a cross normalization constant,
$c_\mathrm{A}$, for the \fpma and \fpmb spectra.  Results for these
fits\footnote{In summary, the \isis syntax for the fitted model
  follows: \texttt{tbabs $\cdot$ (1 $\pm$ constant(1)) $\cdot$
    (thcomp(1, polykerrbb(1)) + relxillCp(1))}, with model
  identifiers, e.g., the \texttt{1} here, assigned based upon
  observation epoch of the dataset being evaluated.} are presented in
Tables~\ref{tab:polykerr_common} and \ref{tab:polykerr_indi}.

% Put Table Comments for Below Table Here:
\newcommand\tcompkerrcom{All fits used $\mbh = 8\,\msun$ and $\dbh =
  16.47$\,kpc.  Likewise, the reflection models used a common powerlaw
  index, $\epsilon_\mathrm{rx}$, for the disk emissivity profiles.
  For \nustar-only fits (first entry), the equivalent neutral column
  was frozen to $\nh = 1.7\times10^{21}\,\cmtwo$. For the joint
  \nicer/\nustar fits (second entry), $\nh$ was modeled, but with
  variable abundances for {O}, {Ne}, and {Fe} ($\mathrm{A_{O}}$,
  $\mathrm{A_{Ne}}$, $\mathrm{A_{Fe}}$) and a variable redshift
  ($z_{\nh}$) in order to account for possible calibration
  uncertainties in the \nicer spectra.  This absorption was modified
  by a dust scattering halo (see text), with the sole fit parameter
  being the distance from observer to halo, relative to the source
  distance ($\xdust$). A number of absorption lines with fixed
  equivalent widths, representing previously detected features (likely
  due to the interstellar medium; see text) were included.  Error bars
  are 90\% confidence level.}  \input{table_polykerr_common}

% Put Table Comments for Below Table Here:
\newcommand\tcompkerrindi{Degrees of freedom for each indivdual set of
  observations includes the global parameters. Three additional
  features (which we use to represent \nicer calibration
  uncertainties) were added to the \nicer spectra: a reverse edge, and
  two absorption lines. (See text.)  Cross normalization constants
  were added, where we multiply the model by $(1+c_\mathrm{A})$ (\fpma
  spectra), $(1-c_\mathrm{A})$ (\fpmb spectra), or $(1+c_\mathrm{N1})$
  or $(1+c_\mathrm{N2})$ (\nicer spectra).  Error bars are 90\%
  confidence level.} \input{table_polykerr_individual}

Overall this fit is quite successful in describing all ten \nustar
observations with a common set of black hole parameters.  The spin and
disk inclination, commensurate with the suggestion by
\citet{parker:19a}, are in fact formally strongly
constrained\footnote{The MCMC derived probability distribution for
  $a^*$ is narrow enough such that the function of $p$ that would
  yield a uniform prior on $a^*$ does not vary strongly over that
  posterior range.}. The disk accretion rates, $\dot{M}_\mathrm{dd}$,
and color-correction factors, $\fc$, vary for the individual
observations. We have verified with various spot checks that the fits
are in fact degenerate with different masses, distances, accretion
rates, and color-correction factors scaling as suggested from the
discussion above.  Only the spin and inclination do not change among
these degenerate fits, with lower spins and/or lower inclinations
formally being strongly disfavored for this assumed model.  The
relative trends of accretion rate and disk color-correction factor
remain consistent, however, even if their absolute values change based
upon assumed mass and distance.  The hard tail generally increases in
strength with increasing accretion rate, but the behavior is more
complex than this simple statement.  We discuss these issues further
in \S\ref{sec:discuss}.

\begin{figure}
\begin{center}
\includegraphics[width=0.46\textwidth,viewport=85 20 580 380]{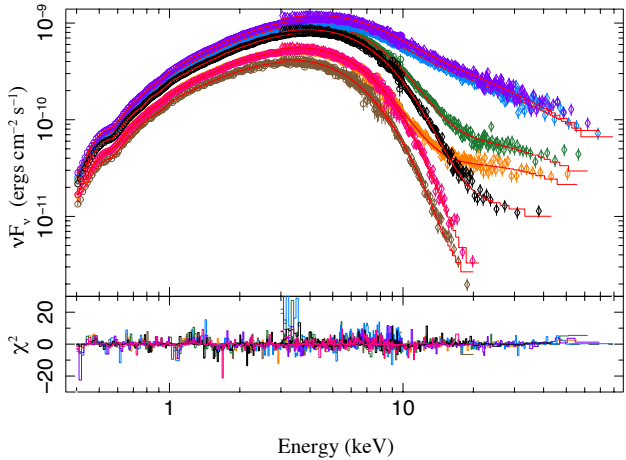}
\end{center}
\caption{Flux-corrected spectra and fits using Comptonized (\thcomp)
  disk atmosphere (\polykerrbb) models jointly fit to \nicer/\nustar
  spectra of \fu.  The \nustar-\fpma and \fpmb modules were fit
  separately, but are combined for presentation (likewise for
  individual \nicer ObsIDs that make up an individual observing
  epoch.)} \label{fig:joint_spectra}
\end{figure}

\subsection{\nicer/\nustar Joint Spectral Fits} \label{sec:nicernustar}

There are seven epochs of observations with joint \nicer/\nustar
spectra.  As shown in Table~\ref{tab:obs_log}, there is good overlap
between the \nicer and \nustar observing windows, and as discussed in
\S\ref{sec:timing} there is only a small amount of variability in the
X-ray lightcurve on any time scale covered by a single observing
epoch.  We therefore fit the \nicer and \nustar spectra together.
\nicer, however, covers a lower energy bandpass, and thus we have to
include more complexity in the absorption model.  Additionally, \nicer
is a somewhat newer instrument that is subject to greater calibration
uncertainties.  We account for both of these facts in our fits.  For
these joint fits we present a Comptonized disk model exactly as
discussed above, with the following changes to account for the \nicer
data.

As for the \fpma and \fpmb spectra, we introduce cross normalization
constants, $c_\mathrm{N}$, for the \nicer spectra. Multiplying the
model by $(1+c_\mathrm{N})$ essentially normalizes the \nicer fit to
the average of the \fpma and \fpmb fits.  Since the field of view of
\nicer is comparable to the typical size of a dust scattering halo, we
apply the \xscat dust model \citep{smith:16c} to all spectra.  We tie
the neutral column to the value from the \tbabs model (which we now
allow to be a free parameter, instead of being fixed).  We choose the
MRN dust model for the scattering. We choose the radii of the circular
extraction regions as 50\,\arcsec\ for the \nustar spectra and
180\,\arcsec\ for the \nicer spectra.  The lone fit parameter for the
\xscat dust model then becomes the distance of the halo from the
observer, relative to the distance of the source, $\xdust$.  This
parameter was fit to be the same for all spectra.

\begin{figure*}
\begin{center}
\includegraphics[width=0.46\textwidth,viewport=85 35 580 525]{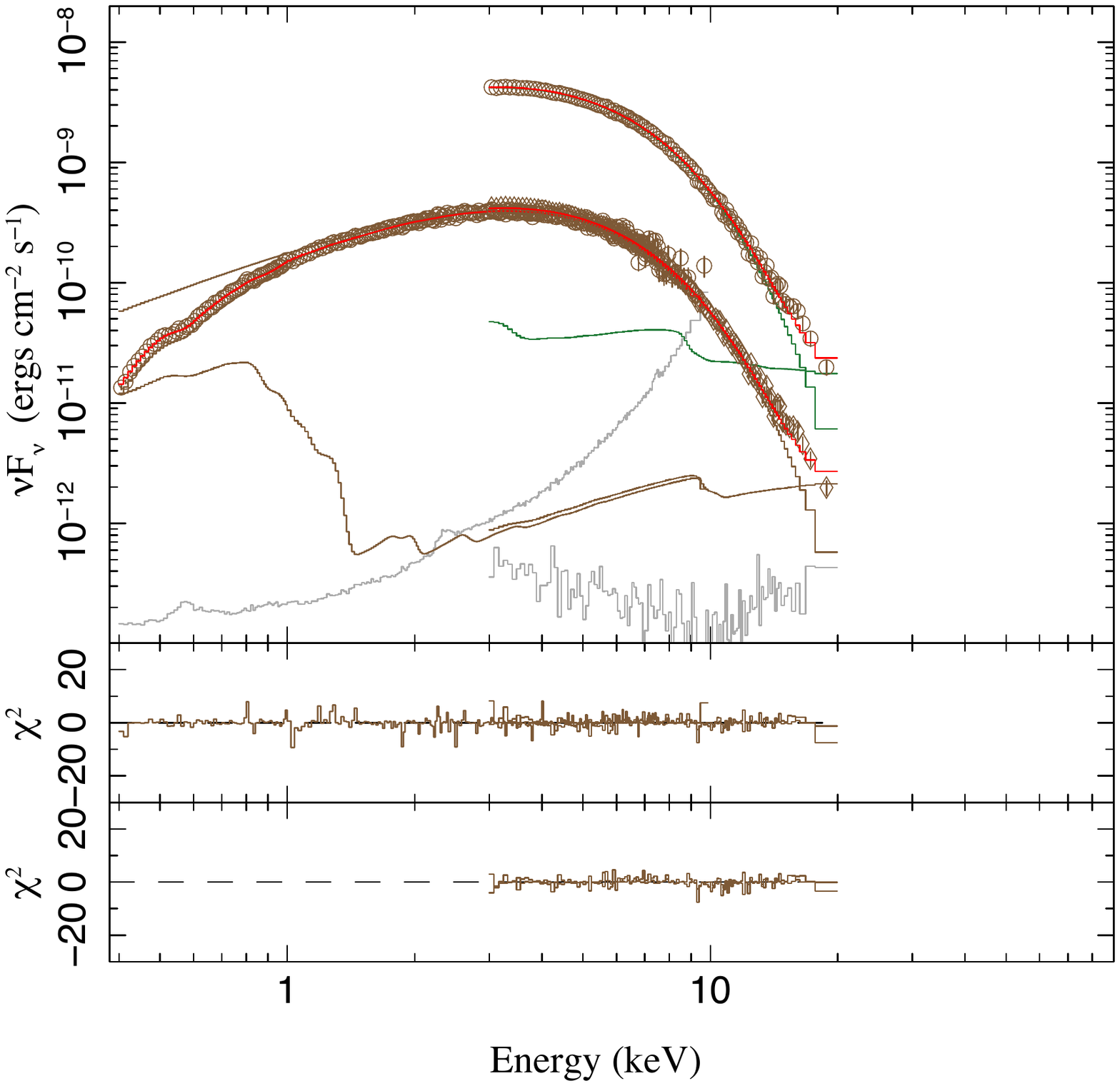}
\includegraphics[width=0.46\textwidth,viewport=85 35 580 525]{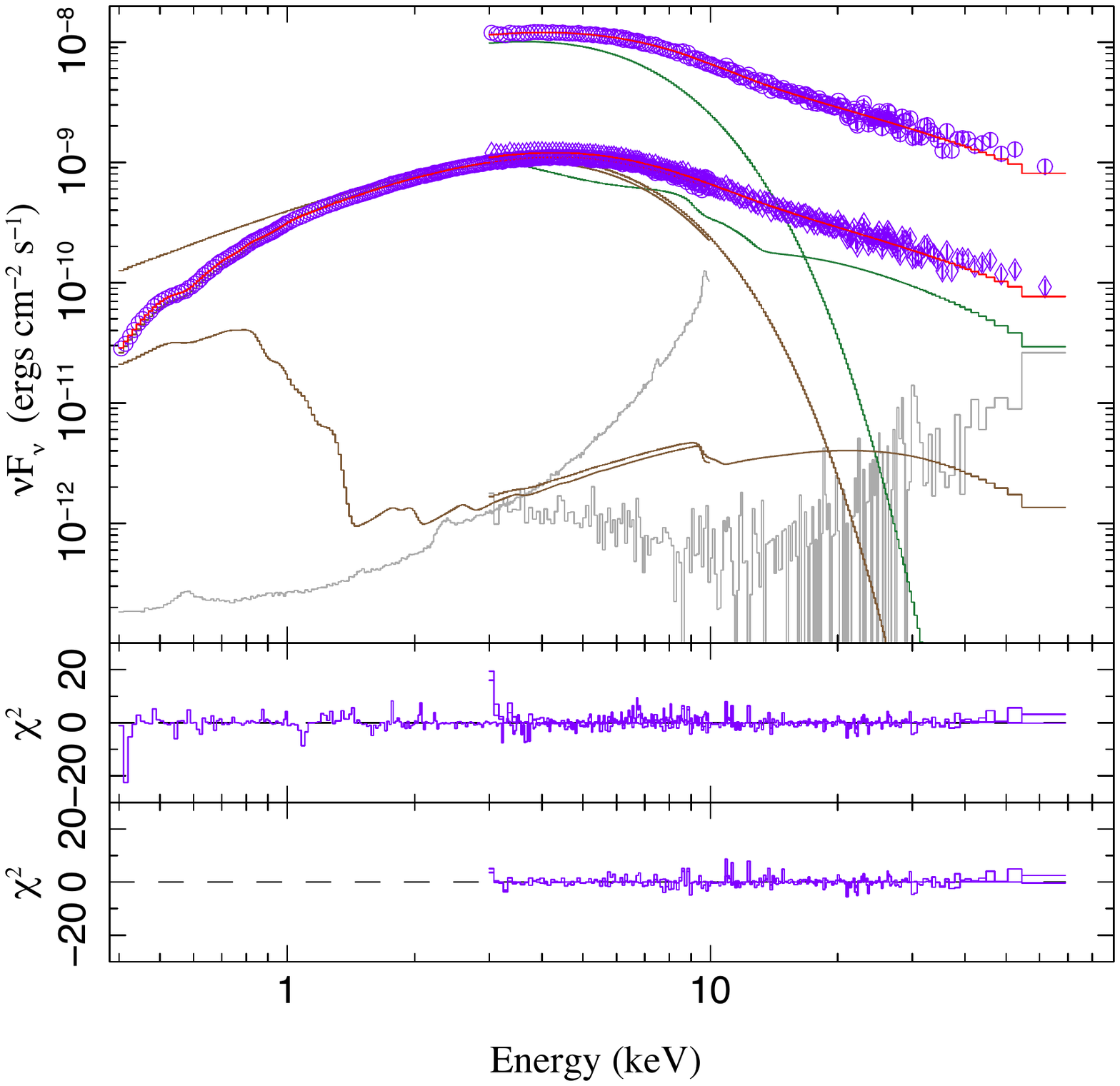}
\end{center}
\caption{Flux-corrected spectra from faint (epoch 2) and bright (epoch
  8) periods, fit with Comptonized (\thcomp) disk atmosphere
  (\polykerrbb) models see Tables~\ref{tab:polykerr_common} and
  \ref{tab:polykerr_indi}).  Here we show individual model components.
  For the joint \nicer/\nustar spectra, grey lines show the background
  components and brown lines show the \relxillCp components (without
  absorption or other line/edge features) as well as the \polykerrbb
  disk component, absent Comptonization, both with and without
  absorption/line/edge features included. The \nustar-only spectra,
  along with the same model components shown in green, are offset
  upward by a factor of ten.} \label{fig:component_spectra}
\end{figure*}

Adding the dust model and a variable absorption to the model still
leaves a number of residuals in the soft end ($\lesssim 2$\,keV) of
the \nicer spectra.  These could be a combination of: unmodeled
interstellar medium (ISM) features, systematics in our assumed model
(e.g., the assumption of an energy-independent color-correction
factor, $\fc$ for each disk spectrum, or the details of the soft
excess associated with the \relxillCp component), and/or calibration
uncertainties/errors in the \nicer response function.  For ISM
features, we add several fixed equivalent width absorption features
using \gabs models with fixed energies (0.530\,keV, 0.654\,keV,
0.666\,keV, 0.849\,keV, 0.855\,keV, 0.922\,keV), widths (0.01\,keV),
and strengths to represent ISM features previously observed with the
\chandra-High Energy Transmission Gratings in observations of \fu
\citep{nowak:11b}.  This still left residuals in the soft X-ray
energies, some of which were alleviated by allowing the \tbabs O, Fe,
and Ne abundances and overall redshift, $z_{\nh}$, to be free
parameters.  We deem these to most likely represent calibration errors
in the \nicer responses.  Several additional features, likely due to
calibration effects, were addressed by adding a `reverse edge' at
$\approx 1$\,keV (with optical depth $\approx 0.04$), and two further
\gabs absorption lines, again with widths fixed to 0.01\,keV and
energies of $\approx 0.572$\,keV and $\approx 0.763$\,keV.  It should
be noted that all of these features, although statistically
significant owing to the high count rates of the spectra, represented
only a few percent deviations from the baseline model.

Fitting this modified Comptonized, relativistic disk model yielded
good results, as seen from Tables~\ref{tab:polykerr_common}
and~\ref{tab:polykerr_indi}, and Figure~\ref{fig:joint_spectra}.
Parameter trends for these fits are very similar to those for the
\nustar-only fits, but we find that both the black hole spin and disk
inclination are systematically higher, while the fitted disk
color-correction factors are systematically lower.

These systematic differences between the \nicer and \nustar spectra
are most evident in the fit residuals and model components shown in
Figures~\ref{fig:joint_spectra} and \ref{fig:component_spectra}. The
higher color-correction factor of the \nustar-only fits is apparent
from the steeper high energy drop of the disk component. Likewise we
see that the \nustar-only fits yield a steeper reflection component,
although in both cases this component is weak in the Fe~K band.  In
fact, in these models reflection if truly present is strongest in the
$\lesssim 1$\,keV \nicer spectra, comprising $\lesssim 20\%$ of the
flux in the 0.4--0.8\,keV energy band.  This soft component of the
reflection model, however, is not responsible for the differences in
fitted inclination and color-correction between the joint
\nicer/\nustar fits and the \nustar-only fits.  If we restrict the
\nicer spectra to the 3--10\,keV band, the former fits still yield the
same higher inclination and lower color-correction values when
compared to the \nustar-only fits.

If one instead postulates a `slope' difference between the \nicer and
\nustar calibrations, then altering the \nicer spectra effectively by
a $|\Delta \Gamma| \approx 0.03$--0.07 greatly improves the fit
($\Delta \chi^2 \approx 10^3$, when including the 0.4--10\,keV \nicer
spectra), and drives the fitted spin, inclination, and
color-correction factors closer towards the \nustar-only fitted
values.  (We do not present results from these latter fits in the
remainder of this work.)  We have not found any other straightforward
model that improves the agreement between the \nicer and \nustar
spectra, and hypothesize that the remaining differences are dominated
by systematic calibration effects.

\section{Discussion}\label{sec:discuss}

\nustar detects the faintest spectra of \fu all the way out to
20\,keV, yet these spectra are nearly completely dominated by a disk
spectrum with virtually no contribution from a coronal component.
This is true for both the \eqpair and Comptonized disk fits, with the
fraction of scattered photons being less than half a percent (see
Tables~\ref{tab:eqp} and~\ref{tab:polykerr_indi}).  At the opposite
extreme, the brightest observations indicate significant scattering of
the underlying disk spectra with optical depths reaching $\taues
\approx 0.4$ (\eqpair models) and covering fractions reaching
${\approx}75\%$ (Comptonized disk models).

\begin{figure*}
\begin{center}
\includegraphics[width=0.32\textwidth,viewport=85 20 580 380]{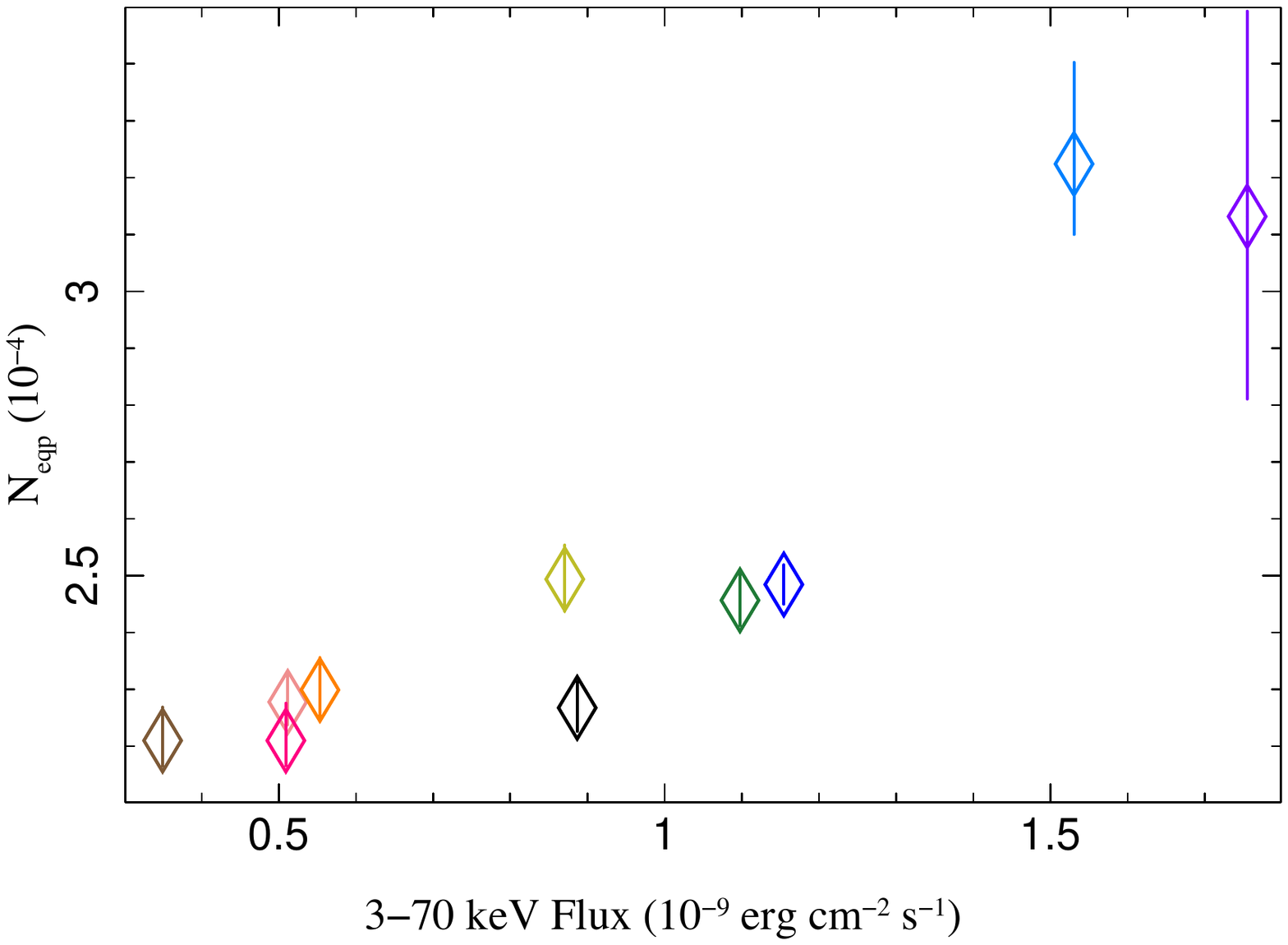}
\includegraphics[width=0.32\textwidth,viewport=85 20 580 380]{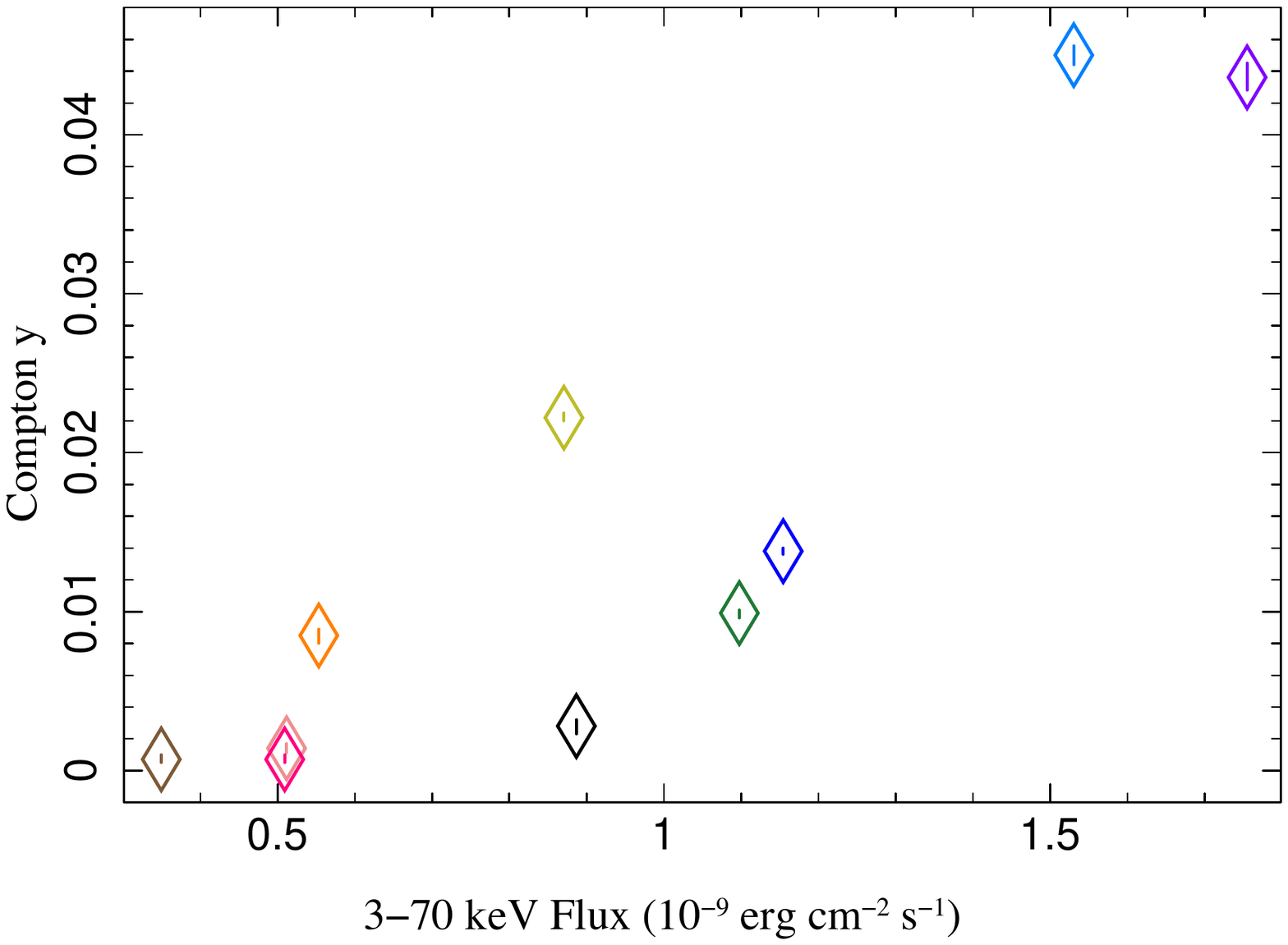}
\includegraphics[width=0.32\textwidth,viewport=85 20 580 380]{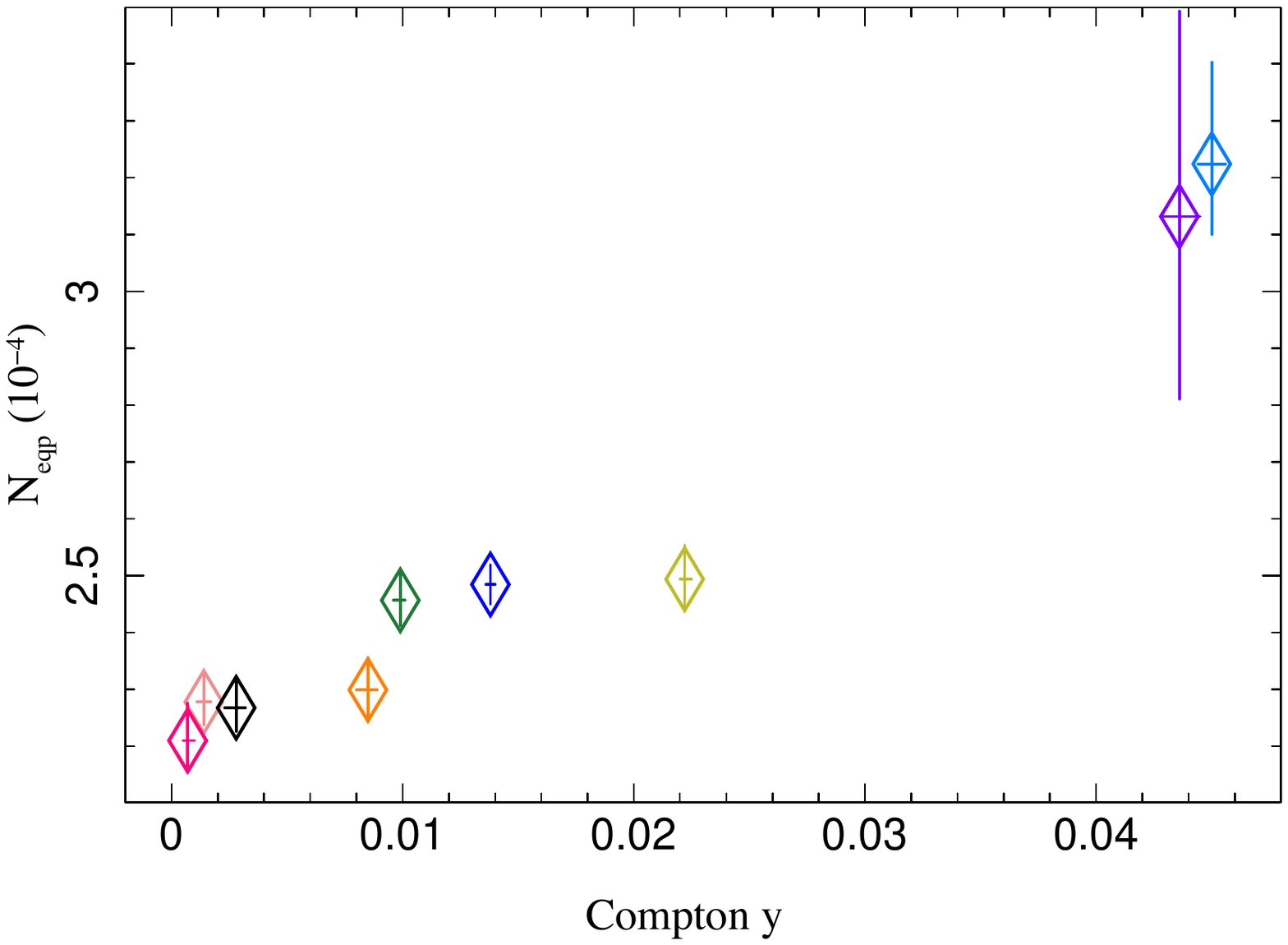}
\end{center}
\caption{Correlations among parameters from \eqpair model fits to the
  ten \nustar-only spectra.  All parameters represent the best fit
  values plus 90\% confidence interval around the median, obtained
  from MCMC analyses of our best fit models.  \textsl{Left:} Model
  normalization vs.\ absorbed 3--70\,keV flux.  \textsl{Middle:}
  Derived Compton $y$ parameter vs.\ 3--70\,keV flux.  Our fits always
  have optical depth $\taues < 1$, so $y=4 kT_e/m_e c^2 \taues$.
  \textsl{Right:} Model normalization vs.\ Compton $y$
  parameter. }\label{fig:nustar_neqp_v_flux}
\end{figure*}

Generally, the strength of the powerlaw tail increases with overall
flux, but there is not a single track.  With the ten \nustar
observations discussed here, there appears to be at least two tracks,
both showing a hard tail increasing with rising flux, but with one
having significantly stronger tails.  These two tracks are quite
apparent in the \eqpair fits when plotting the Compton $y$ value
vs.\ observed flux, as shown in
Figure~\ref{fig:nustar_neqp_v_flux}. Comparing these two tracks in
terms of the long term \maxi lightcurve presented in
Figure~\ref{fig:maxi}, it is difficult to discern a pattern related to
the \fu spectra being on one track versus the other.

Multiple tracks are also discernible when comparing \eqpair
normalization to observed flux, with the lowest normalizations
occurring at the lowest flux levels.  Two track behavior becomes less
obvious when comparing \eqpair normalization to Compton $y$ parameter,
with the lowest normalizations being associated with the smallest $y$
parameters. The \eqpair normalization scales as
\begin{equation}
\neqp \propto \rin^2 D^{-2} \fc^{-4} \cos i ~~,
\end{equation}
where $\rin$ is the inner disk radius (and therefore also scales with
black hole mass).  The increase in normalization with increasing flux
therefore can be related to an increase in the inner disk radius, a
decrease in color correction factor, or a decrease in the inclination.
Although changes in inclination are possible via, for example, disk
warping (e.g., \citealt{pringle:96a}), it would be highly unexpected
to be so tightly correlated with flux, especially on these time
scales.  We argue below that color-correction factor changes provide
the most compelling explanation.

\begin{figure}
\begin{center}
\includegraphics[width=0.45\textwidth,viewport=65 22 580 522]{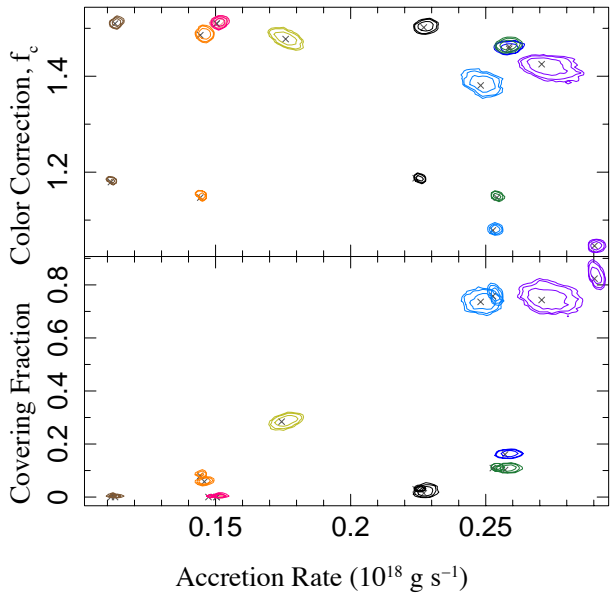}
\end{center}
\caption{Correlations from \thcomp{}$\otimes$\polykerrbb model fits to
  the \nustar-only spectra and \nicer/\nustar spectra.  We show
  68\%/90\%/95\% confidence intervals obtained from MCMC analyses of
  our best fits.  \textsl{Top panel:} Disk color-correction
  vs.\ accretion rate. The \nustar-only spectra yield the higher
  color-correction factors. \textsl{Bottom panel:} Compton corona
  covering fraction vs.\ disk accretion rate. The \nustar-only spectra
  yield the broader confidence contours.}\label{fig:polykerr_v_rate}
\end{figure}

\begin{figure}
\begin{center}
\includegraphics[width=0.45\textwidth,viewport=85 20 580 380]{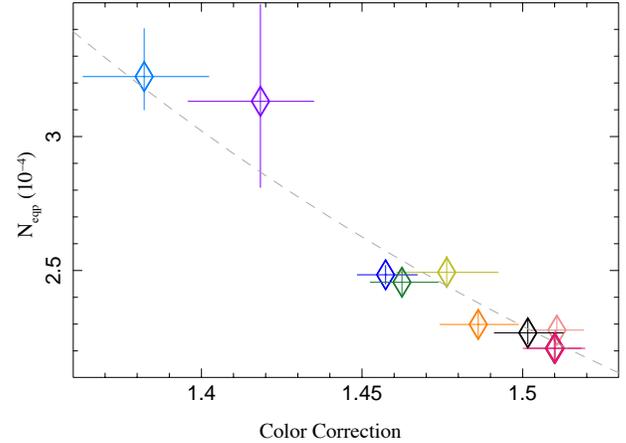}
\end{center}
\caption{Model normalization for \eqpair model vs.\ color-correction
  factor for Comptonized (\thcomp) disk atmosphere (\polykerrbb) model
  fits to \nustar-only spectra of \fu. The dashed line shows scaling
  the normalization $\propto \fc^{-4}$.}\label{fig:norm_vs_color}
\end{figure}

The two track behavior is also apparent in the Comptonized,
relativistic disk fits, as shown in
Figure~\ref{fig:polykerr_v_rate}. Here we show the corona covering
fractions and disk color-correction factors versus disk accretion
rate.  The observations with more prominent hard tails, for both
\nustar-only and \nicer/\nustar spectral fits, have higher coronal
covering factors and lower color-correction factors.  We can correlate
this behavior with \eqpair fits.  The \diskpn model that underlies the
seed photon distribution for the \eqpair model is still somewhat
phenomenological compared to the \kerrbb model at the heart of our
\thcomp{}$\otimes$\polykerrbb fits. One can imagine that the increase
in \eqpair normalization is in fact associated with a true increase in
disk radius as opposed to a decrease in color-correction factor.  The
\kerrbb model, on the other hand, has a fixed inner radius in
geometrical units of $G\mbh/c^2$, and attempts to fit the black hole
spin, disk inclination and accretion rate based upon both the energy
of the spectral peak and the breadth of this peak. It is the breadth
of this peak, more precisely measured with these \nustar spectra than
for any previous observations of \fu, that leads to such a strong
constraint on fitted spin and disk inclination as these spectra are in
the least degenerate portion of parameter space for our assumed model
\citep{parker:19a}.  Correlating the \eqpair normalizations with the
Comptonized relativistic disk color-corrections, we see in
Figure~\ref{fig:norm_vs_color} that the normalizations follow the
$\fc^{-4}$ behavior expected if only this latter parameter, and not
the disk inner radius, varies.  The fact that the
\thcomp{}$\otimes$\polykerrbb fits, which again postulate an
unchanging inner disk radius, provide excellent fits at least in the
\nustar-only case argues strongly for color-correction changes
dominating.

Disk color-corrections are essentially due to electron scattering in
the upper atmosphere of the disk \citep{davis:05a}.  We thus are drawn
to a scenario where as electron scattering is increased in a hot
corona it is decreased in the upper atmosphere of the optically thick
accretion disk itself.  What are we to then make of the systematic
difference between the fitted color-correction factor for the
\nustar-only and \nicer/\nustar fits?  We hypothesize that this is
predominantly a systematic effect owing to calibration error in one or
both of \nustar and \nicer. Spin and inclination estimates, especially
at these high fitted values, are strongly driven by the \emph{breadth}
of the spectral peak.  A high spin, near edge-on disk has a very broad
spectral peak since we are seeing both extreme red and blueshifts of
the disk spectrum.  The \nicer spectra, as discussed above for the
\eqpair fits, show a spectral hardening compared to \nustar of
$|\Delta \Gamma| \approx 0.03$--0.07.  The peak of the \nicer spectra
is \emph{broader} than that seen by \nustar, and therefore the disk
must be more highly inclined to fit the spectrum.  The peak of the
spectrum, however, is not changed (leading to a similar accretion rate
for a given assumed mass) and the overall flux must remain unchanged,
therefore the color-correction factor must drop to compensate for the
higher inclination.  The ratio of fitted color-correction factors
between the two sets of fits is in fact almost exactly given by the
ratio of the cosine of the fitted disk inclination angles as expected
for this systematic dependence.  

Of the two fitted values, the \nustar-only value of $i\approx
75^\circ$ is more plausible, as the \nicer/\nustar value of $i \approx
84^\circ$ should have led to discernible eclipsing of the disk.  A
$75^\circ$ inclination will not lead to an eclipse of the primary by
the Roche lobe of the secondary if the primary mass is $\gtrsim
2\,\msun$ (for a $0.5\,\msun$ secondary) or is $\gtrsim 4\,\msun$ (for
a $1\,\msun$ secondary).  These values increase to $\gtrsim 50\,\msun$
and $\gtrsim 160\,\msun$, respectively, for an $85^\circ$ inclination.
This is ignoring any warp or raised edge/atmosphere in the outer disk,
which could lead to quasi-periodic obscuration at lower inclination
angles, as has been observed in the so-called ``dipping'' sources
(e.g., 4U~1624$-$490; \citealt{xiang:07a}).  It is debatable whether
or not dipping should be observed in \fu if the inclination is
$\approx 75^\circ$.  Although dipping has been observed in systems
with likely lower inclination angles, its presence or absence can be
spectrally-dependent (e.g., XB~1254$-$690; \citealt{diaztrigo:09a}).
Furthermore, \citet{galloway:16a} argue that $i\lesssim 75^\circ$ is a
plausible typical inclination angle demarcating the boundary between
sources that are intermittently dipping from those that never exhibit
dipping. We hypothesize that the \nustar spectral calibration is
closer to `accurate' compared to the \nicer spectral calibration,
although we cannot discount the possibility that revisions to both
calibrations would lead to best fits with $i<75^\circ$.

This lack of fit-degeneracies in terms of spin and inclination, but
remaining degeneracies for mass, distance, and color-correction
factor, as well as the variation of \eqpair normalization, is somewhat
counter to the previous \swift \citep{maitra:14a}, \suzaku
\citep{nowak:11b}, and \rxte \citep{nowak:99d} studies of \fu.  We
note, however, that studies with \swift and \suzaku were limited to
energies $\lesssim 10$\,keV, while studies with \rxte had hard X-ray
backgrounds approximately 100 times greater than for these \nustar
spectra, with corresponding lower signal-to-noise in the hard tail.
\nustar is the first instrument to be able to accurately characterize
the hard tail and thereby accurately characterize the breadth of the
disk spectrum peak.  For example, for the faintest \nustar spectra
with the weakest hard tails the nearly pure disk spectrum is measured
out as far as 20\,keV.

We can also ask to what extent our results are driven by systematics
of our model assumptions, and whether or not individual fits support
the same results as our `global model' with fixed black hole spin and
inclination.  The fact that the individually fit \eqpair models yield
the same trends, e.g., for color-correction vs. hard tail strengths,
as our global model gives us optimism that we are detecting robust,
physically meaningful trends.  Likewise, we note that if we relax the
assumption of a common disk inclination for the observations, the
\nustar-only spectral fit improves by only $\Delta \chi^2 \approx 7$
(for 9 additional parameters), with the individual fitted inclinations
varying from the globally fit value by $\lesssim 1^\circ$.

\begin{figure}
\begin{center}
\includegraphics[width=0.45\textwidth,viewport=60 15 585 525]{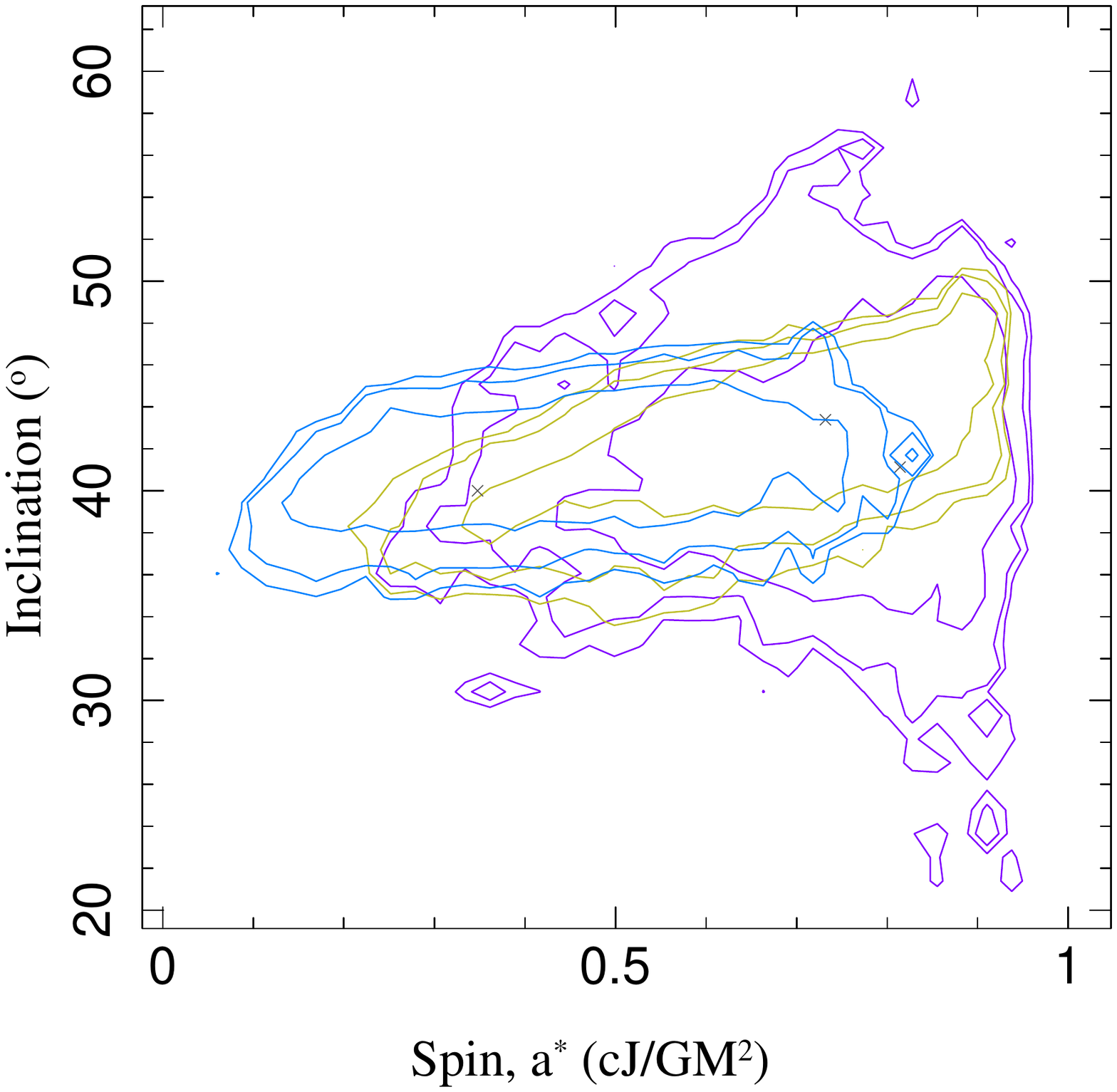}
\includegraphics[width=0.45\textwidth,viewport=60 15 585 525]{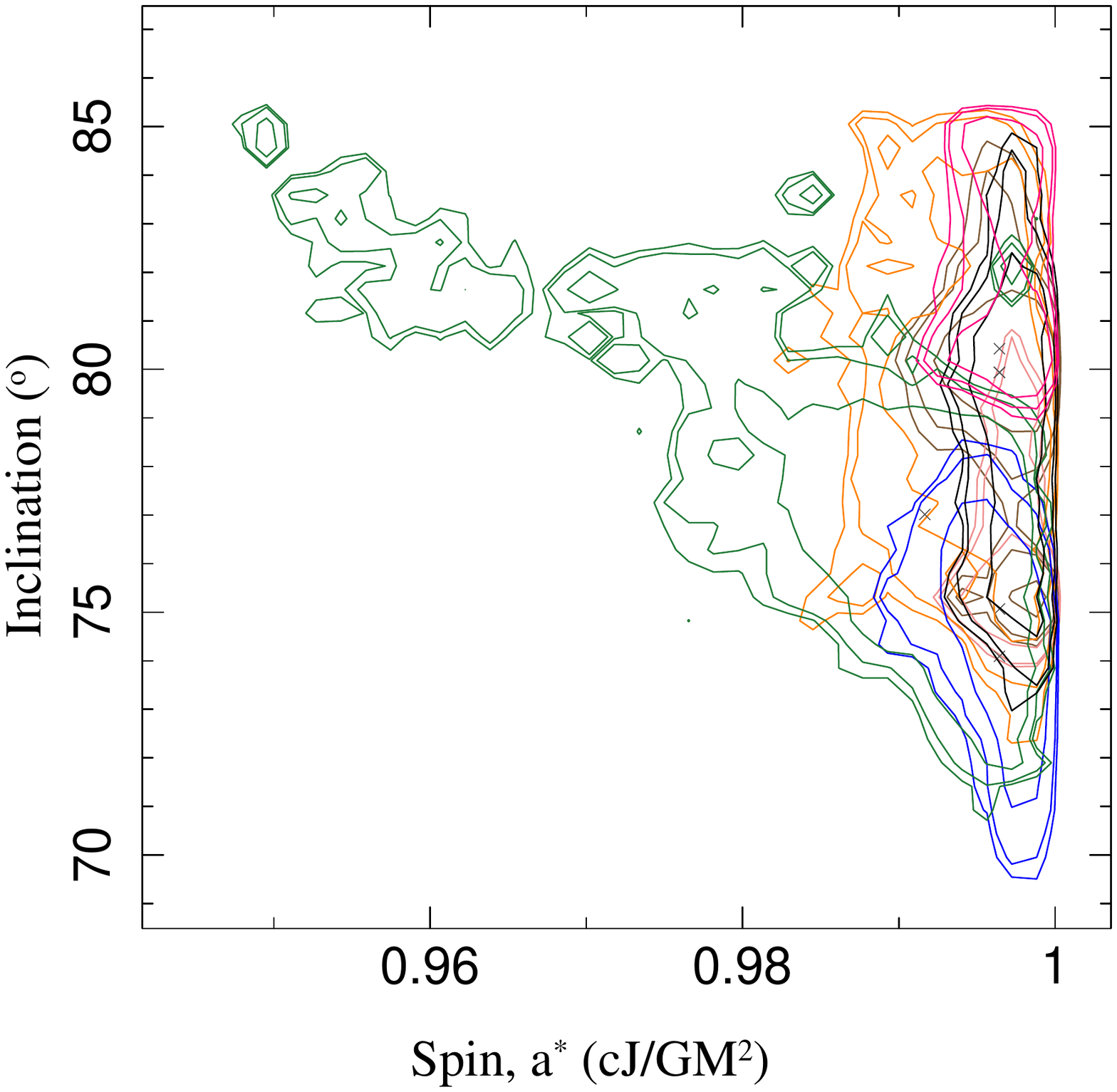}
\end{center}
\caption{Contours of fitted spin vs. inclination, if we allow
  independent spins and inclinations for Comptonized (\thcomp) disk
  atmosphere (\polykerrbb) model fits to \nustar-only spectra of
  \fu. The top figure shows the contours for the three spectra with the
  strongest hard tails, while the bottom panel shows the contours for
  the remaining observations.}\label{fig:spin_vs_inc}
\end{figure}

If we let both spin and inclination be free for each observation, we
find $\Delta \chi^2 \approx 69$ (for 18 additional parameters)
compared to the \nustar-only global model. Almost all of this
improvement comes from epochs 3, 7, and 8, where formally a better fit
is found with disk inclinations $\approx 40^\circ$, spin parameters
ranging from $a^* \approx 0.3$--0.9, and extremely large
color-correction factors $\fc \approx 4.4$--4.8. We show the error
contours (derived from MCMC analyses) for spin vs. inclination for
these fits in Figure~\ref{fig:spin_vs_inc}.  The three observations
that fit lower inclinations and spins exhibit the strongest fitted
coronal components in all of our hypothesized models. The remaining
seven observations have inclination, spin, and color-correction values
comparable to the global fit values.  Despite extensive searches
across parameter space, we have not found any statistically preferred
low inclination, low spin solutions for these latter observations with
weaker fitted coronal components.  We note, however, for the
discussion that follows, that such solutions, if they do exist, would
also require higher fitted color-correction factors. This in turn
would drive the posterior estimates on mass to lower values and the
posterior estimates on distance to higher values.  (See
Figure~\ref{fig:constrain} and the discussion below.)

\begin{figure*}
\begin{center}
\includegraphics[width=0.32\textwidth,viewport=85 20 580 380]{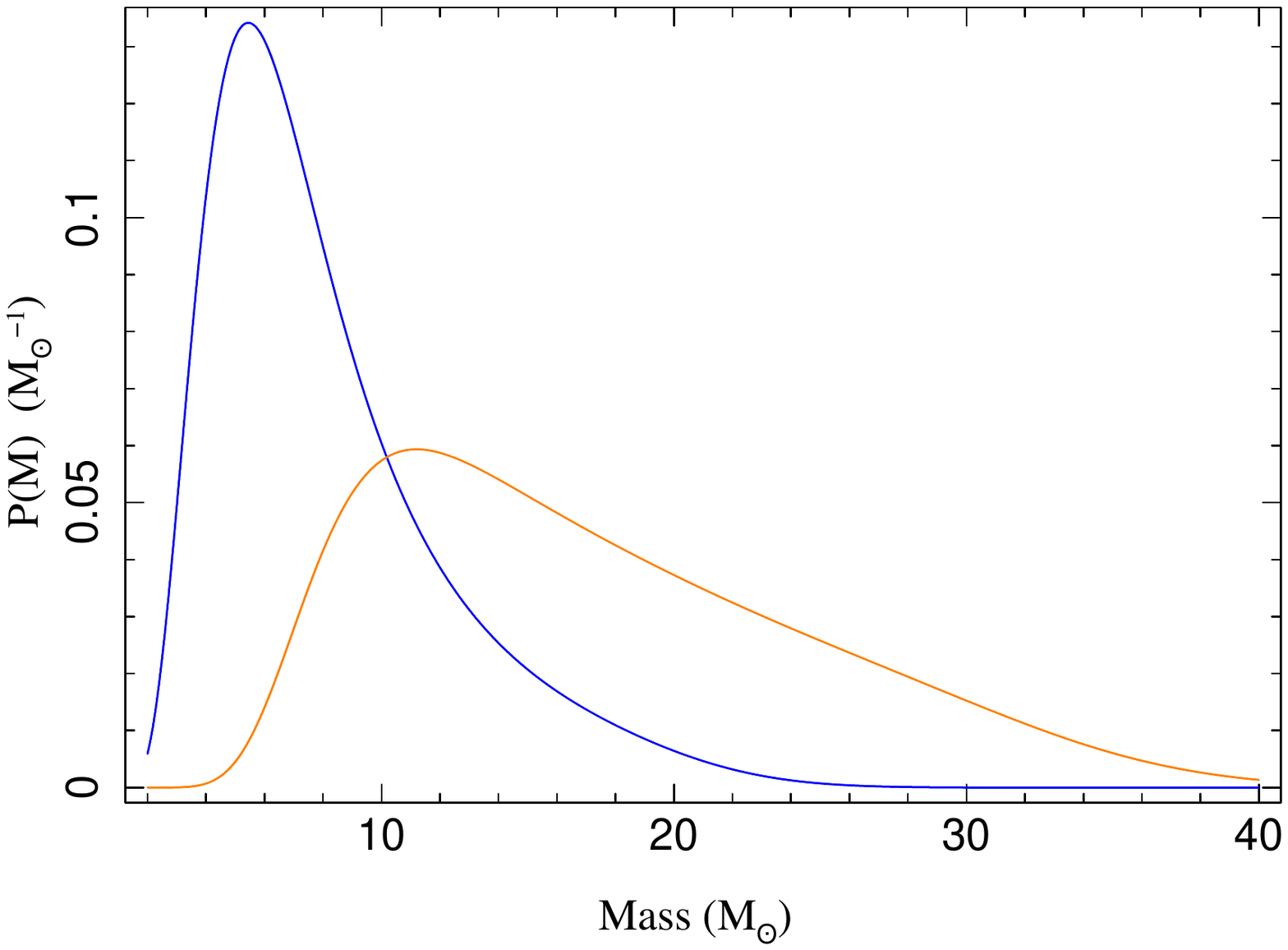}
\includegraphics[width=0.32\textwidth,viewport=85 20 580 380]{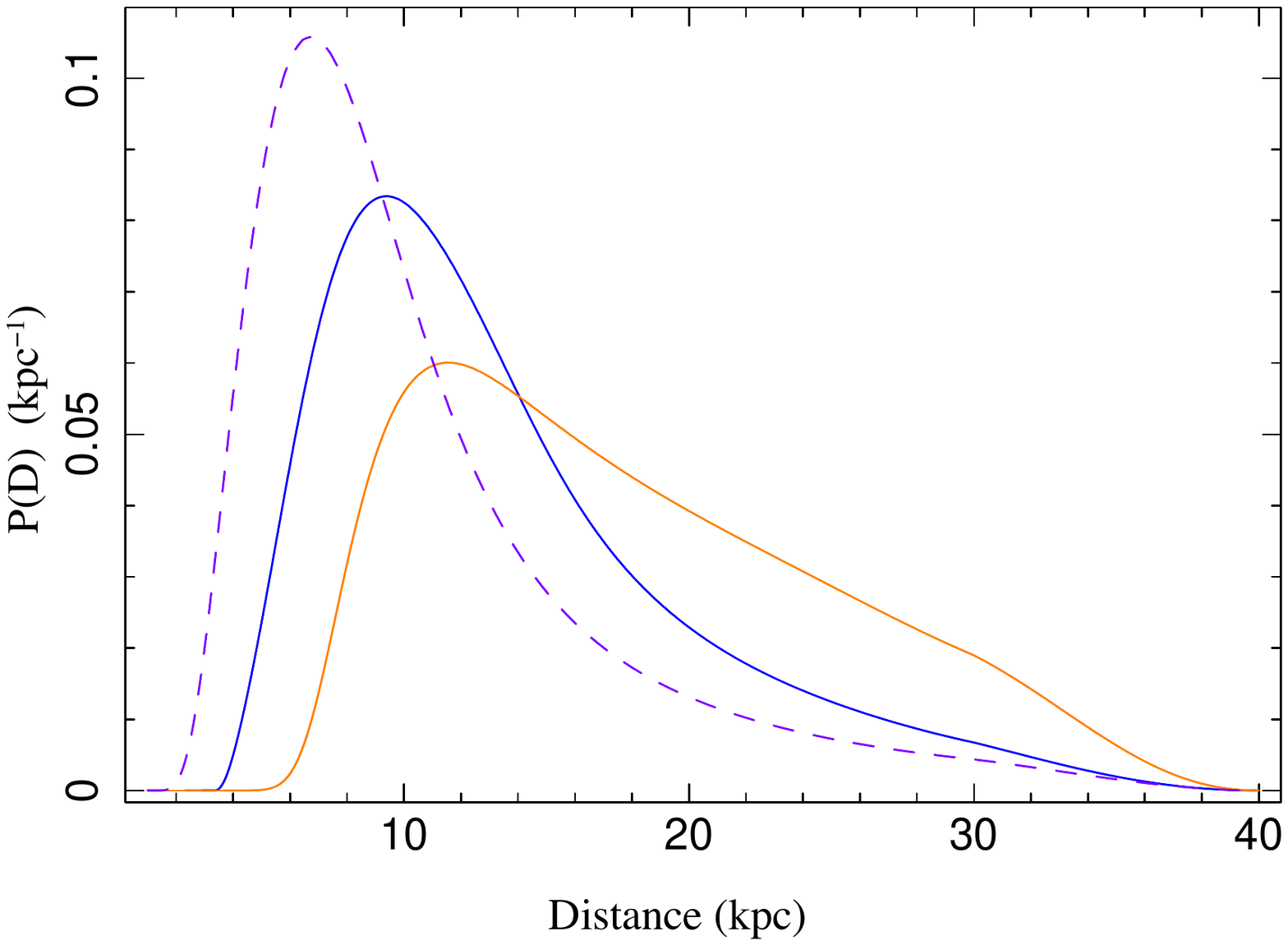}
\includegraphics[width=0.32\textwidth,viewport=85 20 580 380]{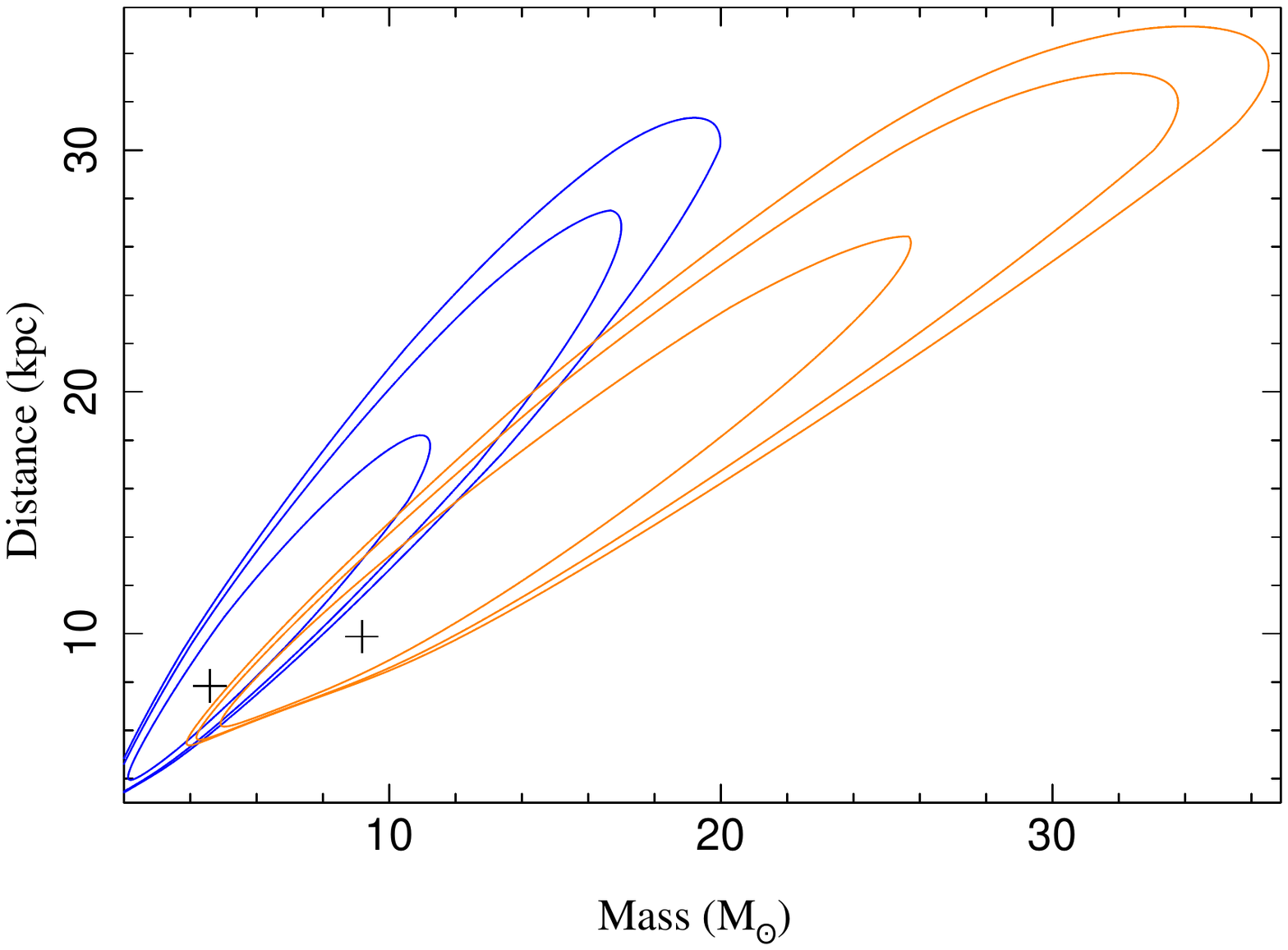}
\end{center}
\caption{Probability distributions for black hole mass (\textsl{Left})
  and distance (\textsl{Middle}) and contours of mass vs.\ distance
  (\textsl{Right}) based upon the \nustar-only fits (blue lines,
  leftmost peaks) and the joint \nicer-\nustar fits (orange lines,
  rightmost peaks) of the flux and color-correction factor for
  Comptonized, relativistic disk models of \fu.  The dashed purple
  line represents the \textsl{Gaia} EDR3 derived probability
  distribution for the distance to the \fu system. As described in the
  text, the other inputs to these curves were assumed distributions
  for the preferred disk atmosphere model color-correction factor and
  fractional Eddington luminosity ratios for \fu.} \label{fig:probs}
\end{figure*}

The statistical uncertainties in our fits are small; uncertainties in
fit parameters are dominated by the systematics of the instrument
calibrations, the fidelity of our assumed model to the physical truth,
and our prior beliefs as to the most `reasonable' parameter values.
Here we use the scaling relations of \S\ref{sec:scale} to combine our
best fit parameters with specific prior probabilities to derive
probability distributions for the mass of and distance to \fu in the
context of our assumed disk model.  We use an updated version
(P. Gandhi 2022, priv.\ comm.) of the \gaia EDR3 distance probability
distribution from \citet{maccarone:20a} which accounted for the fact
that \fu likely lies in the Galactic halo. This probability
distribution peaks at 6\,kpc but still has significant cumulative
probability (17\%) in the 15--30\,kpc range.  For the color-correction
factor, we choose a gaussian prior centered on $\fc = 1.7$ with
$\sigma=0.1$.  Finally, we choose a prior on the fractional Eddington
luminosity for the \emph{faintest} of our observations that is zero at
$L \le 0.005\,\ledd$ and $L \ge 0.08\,\ledd$, linearly rising between
0.005--0.02\,$\ledd$, flat from 0.02--0.04\,$\ledd$, and then linearly
falling between 0.04--0.08\,$\ledd$. For the brightest observation,
multiply these values by approximately five. The concept here is that
if the faintest observation were much fainter (in terms of fractional
Eddington luminosity), and 4U 1957+11 is a black hole, then we would
have observed a transition to the hard state.  If the brightest
observation were much brighter, we likely would have detected greater
X-ray variability, especially given the significant hard tail seen in
our brightest observations.  Combining these distributions with the
scaling relations of \S\ref{sec:scale} and our best fit values for the
Comptonized, relativistic disk fits, we arrive at the probability
distributions shown in Figure~\ref{fig:probs}.

Given these assumptions, for the \nustar-only fits, then the peak of
the mass/distance probability distribution is at $M=4.6\,\msun$ and
$D=7.8$\,kpc, with 50\% of the marginalized mass distribution being at
$M<7.2\,\msun$.  A substantial portion of the mass probability
distribution (22\%) lies within the `mass gap' of $\approx
2$--5\,$\msun$, although higher masses/larger distances do still
easily fit within the posterior probability.  Considering instead a
luminosity/color-correction posterior probability, the peak of the
distribution (for the faintest observation) is at $\lerat \approx
0.017$, which would then imply the brightest observation is at $\lerat
\approx 0.09$.  These values reasonably would allow for a factor of
approximately two fainter state than our faintest \nustar observation
to exist in the \maxi time line, without having exhibited evidence for
a transition to a spectrally hard state.  The marginalized luminosity
distribution (not shown) is essentially zero at values $\lerat
\gtrsim 0.06$ for the faintest observation, which implies an upper
limit of $\lerat \lesssim 0.3$ for the brightest observation.

\section{Summary and Conclusions}

In summary, for these newest X-ray observations of \fu a relativistic
disk, with low absorption column and a modest hard tail, describes the
spectra extremely well.  The lack of any extra components for `surface
emission' and no historical evidence of either bursts or pulsar
periods, has made a neutron star hypothesis less favored.  For the
black hole Comptonized disk models that we have explored, the hard
tail is nearly completely absent in the faintest spectra, yet \nustar
detects the spectra out to 20\,keV.  The tail does increase with
increasing accretion rate, but there are at least two tracks for this
behavior, with one exhibiting a much stronger hard tail than the other
(e.g., compare observations from epochs 4 and 9).  A strong
correlation is seen between the increase of the hard tail (regardless
of track) and decrease of color-correction factor in the disk. None of
these observations exhibit significant short term variability.

The spectral fits strongly constrain the allowed spin and inclination
of \fu given our assumed disk model.  Uncertainties in these
parameters are almost wholly systematic in nature and are related to
the fidelity with which our assumed model describes the underlying
physical truth as well as the accuracy of the instrumental
calibrations. These latter concerns mostly affect the fitted
inclination, which in turn systematically alters the disk
color-correction factor.  All models that we have explored wherein we
apply a uniform spin for all observations yield best fit disk spectra
for a near maximally spinning black hole.  

Allowing spins and inclinations to vary among fits to individual
epochs only modestly improves the statistics of the fits with only
three of the observations fitting lower inclinations and spins, but
with substantially higher color-correction factors.  These three
spectra also have the largest contributions from coronal components in
the context of our assumed models.  If these solutions are more
indicative of the underlying true system parameters, decreasing the
fitted color-correction fractions while ensuring that the implied
luminosities do not fall within a low fractional Eddington range that
likely would have led to an observed spectrally hard state, would push
the \fu mass lower and distance higher in our posterior estimates.
That is, an extremely low mass black hole (or high mass neutron star)
would become more likely. This remains perhaps one of the greater
concerns in our modeling, i.e., that we have failed to discover a
Comptonization scenario that mimics the high spin/high inclination
fits of our disk-dominated models.

That being said, we find the \nustar-only fits to provide a believable
inclination solution, consistent with the lack of any observed
eclipses by the Roche lobe of the secondary, and plausibly consistent
with the lack of dips (but on the cusp of where such dips might be 
expected to occur). The higher inclination found in the
\nicer+\nustar fits would lead to eclipses by the secondary unless the
primary mass were extremely large.  Adding in plausible observational
priors (on fractional Eddington luminosity and distance) plus
theoretical priors (on color-correction factor) constrains the
resulting mass-distance relation to a narrow strip, albeit one that
still covers a wide range of masses and distances.  A large fraction
of the probability distribution, however, resides in the `mass gap' of
2--5\,$\msun$.  If one could confirm the suggested mass ratio of
0.25--0.3 \citep{longa:15a}, then these fits would be consistent with
a black hole of mass $\lesssim 4$\,$\msun$ and distance $< 10$\,kpc.
Improving those optical measurements, however, remains extremely
challenging for the \fu system.

\begin{acknowledgements}
We thank Poshak Gandhi for providing the \gaia EDR3 probability
distribution for the distance to \fu.  We acknowledge useful
conversations with Tom Maccarone and Jim Buckley, and also with
members of the \nustar X-ray Binaries working group.  We thank the
referee for comments that improved this manuscript.  This research has
made use of \maxi data provided by RIKEN, JAXA and the \maxi
team. Additionally, this research has made use of a collection of ISIS
functions (\texttt{isisscripts}) provided by ECAP/Remeis observatory
and MIT (http://www.sternwarte.uni-erlangen.de/isis/). This work has
been supported by both the \nustar Guest Observer Program via JPL
Contract 1617863 and the \nicer Guest Observer Program via NASA Grant
80NSSC19K1583.  Erin Barillier also gratefully acknowledges the
support of the McDonnell Center for the Space Sciences at Washington
University in Saint Louis and support from the Baines Family Planetary
Science Scholarship, made possible by Dr. Kevin H. Baines, Physics PhD
1982.
\end{acknowledgements}

\facilities{Gaia, MAXI, NICER, NuSTAR}

\software{HEASOFT v6.29c (HEASARC 2021), XSPEC v12.11.1
  \citep{arnaud:96a}, RELXILL v1.2.0 \citep{garcia:14a}, ISIS
  v1.6.2-47 \citep{houck:00a}}

%\bibliographystyle{jwapjbib}
%\bibliography{mnemonic,jw_abbrv,bhc}

\end{document}

%% file: table_nustar_eqpair.tex
\begin{deluxetable*}{rccccccccccr} 
\setlength{\tabcolsep}{0.05in} 
\tabletypesize{\footnotesize} 
\tablewidth{0pt} 
\tablecaption{Parameters for \texttt{eqpair} fits to \nustar spectra.} 
\tablehead{ 
   \colhead{Epoch} & \colhead{${N_\mathrm{eqp}}$} 
   & \colhead{$kT_\mathrm{disk}$} & \colhead{${l_{c}}$} 
   & \colhead{$\tau_\mathrm{seed}$} & \colhead{$y$} 
   & \colhead{$R$}
   & \colhead{${N_{\gamma}}$} & \colhead{$E_{\gamma}$} 
   & \colhead{$\sigma_{\gamma}$} & \colhead{$c_{\mathrm{A}}$} 
   & \colhead{$\chi^2/$DoF} \\ 
   & \colhead{($\times10^{-4}$)} & \colhead{(keV)} 
   & \colhead{$(\times10^{-2})$}
   & \colhead{($\times10^{-2}$)} & \colhead{($\times10^{-2}$)} 
   &
   & \colhead{($\times10^{-4}$)} & \colhead{(keV)} & \colhead{(keV)} 
   & \colhead{($\times10^{-2}$)} } 
\startdata 
   1 
 & \errtwo{2.28}{0.04}{0.04} 
 & \errtwo{1.365}{0.005}{0.007} 
 & \errtwo{0.6}{0.3}{0.1} 
 & \errtwo{1.4}{1.2}{0.6} 
 & \errtwo{0.14}{0.03}{0.03}
 & \errtwo{0.80}{1.05}{0.74}
 & \errtwo{1.3}{0.2}{0.4} 
 & \errtwo{6.80}{0.08}{0.15} 
 & \errtwo{0.3}{0.0}{0.1} 
 & \errtwo{-2.7}{0.1}{0.2} 
     & 303.9/251 \\ 
   2 
 & \errtwo{2.20}{0.07}{0.04} 
 & \errtwo{1.279}{0.004}{0.010} 
 & \errtwo{0.3}{0.3}{0.0} 
 & \errtwo{0.2}{1.2}{0.1} 
 & \errtwo{0.07}{0.03}{0.02} 
 & \errtwo {0.84}{1.02}{0.76}
 & \errtwo{0.6}{0.3}{0.4} 
 & \errtwo{6.90}{0.00}{0.49} 
 & \errtwo{0.3}{0.0}{0.2} 
 & \errtwo{1.9}{0.3}{0.2} 
     & 217.6/223 \\ 
   3 
 & \errtwo{2.51}{0.04}{0.07} 
 & \errtwo{1.410}{0.010}{0.005} 
 & \errtwo{14.7}{0.1}{0.5} 
 & \errtwo{12.3}{0.6}{1.0} 
 & \errtwo{2.22}{0.03}{0.02} 
 & \errtwo{0.06}{0.01}{0.02}
 & \errtwo{0.7}{0.2}{0.6} 
 & \errtwo{6.90}{0.00}{0.58} 
 & \errtwo{0.3}{0.0}{0.3} 
 & \errtwo{0.0}{0.1}{0.2} 
     & 542.6/415 \\ 
   4 
 & \errtwo{2.31}{0.05}{0.07} 
 & \errtwo{1.348}{0.009}{0.006} 
 & \errtwo{7.1}{0.1}{1.4} 
 & \errtwo{2.6}{0.2}{0.4} 
 & \errtwo{0.85}{0.04}{0.05} 
 & \errtwo{0.15}{0.50}{0.14}
 & \errtwo{0.6}{0.5}{0.4} 
 & \errtwo{6.53}{0.15}{0.28} 
 & \errtwo{0.0}{0.3}{0.0} 
 & \errtwo{0.5}{0.3}{0.3} 
     & 366.0/327 \\ 
   5 
 & \errtwo{2.48}{0.04}{0.03} 
 & \errtwo{1.529}{0.004}{0.006} 
 & \errtwo{9.2}{0.1}{0.4} 
 & \errtwo{6.2}{0.4}{0.2} 
 & \errtwo{1.38}{0.02}{0.02} 
 & \errtwo{0.02}{0.08}{0.02}
 & \errtwo{0.9}{0.3}{0.7} 
 & \errtwo{6.90}{0.00}{0.46} 
 & \errtwo{0.3}{0.0}{0.3} 
 & \errtwo{0.5}{0.1}{0.1} 
     & 466.1/417 \\ 
   6 
 & \errtwo{2.47}{0.03}{0.06} 
 & \errtwo{1.528}{0.009}{0.004} 
 & \errtwo{6.7}{0.1}{0.5} 
 & \errtwo{4.3}{0.2}{0.6} 
 & \errtwo{0.99}{0.02}{0.03} 
 & \errtwo{0.05}{0.17}{0.05}
 & \errtwo{0.6}{0.6}{0.5} 
 & \errtwo{6.46}{0.29}{0.23} 
 & \errtwo{0.1}{0.2}{0.1} 
 & \errtwo{1.3}{0.2}{0.2} 
     & 373.9/359 \\ 
   7 
 & \errtwo{3.21}{0.19}{0.11} 
 & \errtwo{1.434}{0.013}{0.022} 
 & \errtwo{30.7}{0.8}{0.9} 
 & \errtwo{41.7}{4.5}{2.0} 
 & \errtwo{4.50}{0.06}{0.06} 
 & \errtwo{0.02}{0.06}{0.02}
 & \errtwo{1.7}{0.9}{0.6} 
 & \errtwo{6.66}{0.12}{0.10} 
 & \errtwo{0.0}{0.3}{0.0} 
 & \errtwo{0.6}{0.2}{0.1} 
     & 430.7/435 \\ 
   8 
 & \errtwo{3.13}{0.36}{0.32} 
 & \errtwo{1.491}{0.022}{0.024} 
 & \errtwo{29.2}{0.9}{1.3} 
 & \errtwo{42.8}{5.7}{4.2} 
 & \errtwo{4.36}{0.09}{0.08} 
 & \errtwo{0.02}{0.07}{0.02}
 & \errtwo{0.3}{1.2}{0.3} 
 & \errtwo{6.66}{0.20}{0.42} 
 & \errtwo{0.0}{0.3}{0.0} 
 & \errtwo{1.1}{9.2}{8.9} 
     & 423.1/400 \\ 
   9 
 & \errtwo{2.29}{0.02}{0.06} 
 & \errtwo{1.512}{0.011}{0.003} 
 & \errtwo{2.0}{0.0}{0.8} 
 & \errtwo{1.7}{0.3}{0.7} 
 & \errtwo{0.28}{0.04}{0.05} 
 & \errtwo{0.57}{1.17}{0.53}
 & \errtwo{1.8}{0.4}{1.2} 
 & \errtwo{6.59}{0.25}{0.29} 
 & \errtwo{0.3}{0.0}{0.2} 
 & \errtwo{0.0}{0.2}{0.1} 
     & 307.1/281 \\ 
   10 
 & \errtwo{2.19}{0.09}{0.02} 
 & \errtwo{1.376}{0.003}{0.012} 
 & \errtwo{0.3}{0.3}{0.0} 
 & \errtwo{0.2}{1.7}{0.1} 
 & \errtwo{0.07}{0.03}{0.02} 
 & \errtwo{0.82}{1.03}{0.74}
 & \errtwo{1.5}{0.5}{0.7} 
 & \errtwo{6.39}{0.19}{0.16} 
 & \errtwo{0.3}{0.0}{0.2} 
 & \errtwo{-0.9}{0.3}{0.2} 
     & 256.0/225 \\ 
All & & & & & & & & & & & 3686.9/3333 \\ 
\enddata 
\tablecomments{\tcomeqp} \label{tab:eqp} 
\end{deluxetable*}

%% file: table_polykerr_common.tex
\begin{deluxetable*}{cccccccccr} 
\setlength{\tabcolsep}{0.05in} 
\tabletypesize{\footnotesize} 
\tablewidth{0pt} 
\tablecaption{Global parameters for Comptonized disk fits to \nustar and joint \nicer/\nustar spectra.} 
\tablehead{ 
%   \colhead{Observatory} 
     \colhead{$a^*$} & \colhead{$ i $} 
   & \colhead{$\mathrm{N_{H}}$} & \colhead{$\mathrm{A_{O}}$} 
   & \colhead{$\mathrm{A_{Ne}}$} & \colhead{$\mathrm{A_{Fe}}$} 
   & \colhead{$z_{\mathrm{N_H}}$} & \colhead{$x_{\mathrm{dust}}$} 
   & \colhead{$\epsilon_\mathrm{rx}$} 
   & \colhead{$\chi^2/$DoF}  \\ 
   & \colhead{($^\circ$)} 
   & \colhead{($\times 10^{21}\,\mathrm{cm^{-2}}$)}  
   &  &  &  &  &  \colhead{($\times10^{-3}$)} } 
\startdata 
% \nustar 
   \errtwo{0.9961}{0.0003}{0.0003} 
 & \errtwo{75.00}{0.13}{0.28} 
   & \textsl{1.7} & \nodata & \nodata 
   & \nodata & \nodata & \nodata 
 & \errtwo{5.00}{0.00}{0.09} 
   & 3569.1/3351 \\ 
% Both 
   \errtwo{0.9980}{0.0000}{0.0001} 
 & \errtwo{84.05}{0.15}{0.18} 
 & \errtwo{0.95}{0.01}{0.02} 
 & \errtwo{1.3}{0.0}{0.0} 
 & \errtwo{0.0}{0.2}{0.0} 
 & \errtwo{0.3}{0.1}{0.1} 
 & \errtwo{-0.1200}{0.0002}{0.0001} 
 & \errtwo{0}{4}{0} 
 & \errtwo{5.00}{0.00}{0.15} 
   & 7119.3/5244 \\ 
\enddata 
\tablecomments{\tcompkerrcom} \label{tab:polykerr_common} 
\end{deluxetable*}

%% file: table_polykerr_individual.tex
\begin{deluxetable*}{rcccccccccr} 
\setlength{\tabcolsep}{0.05in} 
\tabletypesize{\footnotesize} 
\tablewidth{0pt} 
\tablecaption{Individual observation parameters for Comptonized disk fits to \nustar and joint \nicer/\nustar spectra.} 
\tablehead{ 
   \colhead{Epoch} & \colhead{$f_c$} & \colhead{$\dot M$} 
   & \colhead{$\Gamma_{tc}$} & \colhead{$f_{tc}$} 
   & \colhead{$N_{rx}$} & \colhead{$\log \xi_{rx}$} 
   & \colhead{$c_\mathrm{A}$} & \colhead{$c_\mathrm{N1}$} 
   & \colhead{$c_\mathrm{N2}$} 
   & \colhead{$\chi^2/$DoF}  \\ 
   & & \colhead{($\times10^{18}\,\mathrm{g/s}$)} 
   & & & & & \colhead{($\times10^{-2}$)} 
   & \colhead{($\times10^{-2}$)} & \colhead{($\times10^{-2}$)} } 
\startdata 
 1 
 & \errtwo{1.511}{0.008}{0.008} 
 & \errtwo{0.150}{0.002}{0.002} 
 & \errtwo{3.40}{0.00}{0.02} 
 & \errtwo{0.0}{0.3}{0.0} 
 & \errtwo{3.4}{0.1}{0.9} 
 & \errtwo{2.3}{0.3}{0.4} 
 & \errtwo{-2.8}{0.2}{0.1} 
  & \nodata  & \nodata 
   & 330.0/250 \\ 
 2 
 & \errtwo{1.510}{0.008}{0.009} 
 & \errtwo{0.112}{0.002}{0.001} 
 & \errtwo{3.40}{0.00}{0.13} 
 & \errtwo{0.0}{0.4}{0.0} 
 & \errtwo{1.9}{0.0}{1.1} 
 & \errtwo{2.2}{0.3}{1.1} 
 & \errtwo{1.9}{0.3}{0.2} 
  & \nodata  & \nodata 
   & 227.1/222 \\ 
  \nodata  
 & \errtwo{1.180}{0.006}{0.006} 
 & \errtwo{0.110}{0.001}{0.000} 
 & \errtwo{2.85}{0.17}{0.04} 
 & \errtwo{0.0}{0.8}{0.0} 
 & \errtwo{0.2}{0.0}{0.1} 
 & \errtwo{1.1}{0.2}{0.1} 
 & \errtwo{1.9}{0.3}{0.2} 
 & \errtwo{-6.3}{0.5}{0.6} 
   & \nodata 
   & 579.3/490 \\ 
 3 
 & \errtwo{1.475}{0.017}{0.014} 
 & \errtwo{0.175}{0.005}{0.004} 
 & \errtwo{2.62}{0.03}{0.04} 
 & \errtwo{28.7}{1.8}{2.7} 
 & \errtwo{0.3}{0.1}{0.1} 
 & \errtwo{4.1}{0.3}{0.3} 
 & \errtwo{0.0}{0.1}{0.2} 
  & \nodata  & \nodata 
   & 457.9/414 \\ 
 4 
 & \errtwo{1.487}{0.012}{0.013} 
 & \errtwo{0.145}{0.002}{0.002} 
 & \errtwo{2.26}{0.08}{0.09} 
 & \errtwo{5.5}{1.2}{0.9} 
 & \errtwo{0.1}{0.0}{0.1} 
 & \errtwo{2.7}{0.2}{0.5} 
 & \errtwo{0.5}{0.3}{0.3} 
  & \nodata  & \nodata 
   & 370.3/326 \\ 
  \nodata  
 & \errtwo{1.148}{0.008}{0.006} 
 & \errtwo{0.144}{0.001}{0.001} 
 & \errtwo{2.44}{0.04}{0.05} 
 & \errtwo{8.3}{0.9}{0.8} 
 & \errtwo{0.2}{0.0}{0.1} 
 & \errtwo{0.8}{0.1}{0.1} 
 & \errtwo{0.5}{0.3}{0.3} 
 & \errtwo{-6.8}{0.4}{0.4} 
 & \errtwo{-6.9}{0.4}{0.4} 
   & 984.9/912 \\ 
 5 
 & \errtwo{1.458}{0.009}{0.009} 
 & \errtwo{0.258}{0.003}{0.004} 
 & \errtwo{2.52}{0.05}{0.04} 
 & \errtwo{15.8}{1.3}{1.0} 
 & \errtwo{0.1}{0.2}{0.1} 
 & \errtwo{2.7}{0.7}{1.3} 
 & \errtwo{0.5}{0.1}{0.1} 
  & \nodata  & \nodata 
   & 426.1/416 \\ 
 6 
 & \errtwo{1.462}{0.012}{0.009} 
 & \errtwo{0.258}{0.003}{0.004} 
 & \errtwo{2.41}{0.08}{0.06} 
 & \errtwo{10.4}{1.3}{1.0} 
 & \errtwo{0.0}{0.0}{0.0} 
 & \errtwo{4.7}{-0.1}{3.2} 
 & \errtwo{1.3}{0.2}{0.2} 
  & \nodata  & \nodata 
   & 352.2/358 \\ 
  \nodata  
 & \errtwo{1.147}{0.008}{0.006} 
 & \errtwo{0.253}{0.002}{0.001} 
 & \errtwo{2.54}{0.03}{0.05} 
 & \errtwo{11.0}{0.8}{1.1} 
 & \errtwo{0.2}{0.0}{0.0} 
 & \errtwo{0.8}{0.0}{0.1} 
 & \errtwo{1.3}{0.2}{0.2} 
 & \errtwo{-11.2}{0.3}{0.3} 
 & \errtwo{-7.7}{0.3}{0.3} 
   & 1310.6/964 \\ 
 7 
 & \errtwo{1.382}{0.020}{0.019} 
 & \errtwo{0.248}{0.005}{0.005} 
 & \errtwo{2.79}{0.02}{0.03} 
 & \errtwo{73.4}{3.4}{3.3} 
 & \errtwo{1.2}{0.5}{0.7} 
 & \errtwo{2.7}{0.4}{0.5} 
 & \errtwo{0.6}{0.2}{0.2} 
  & \nodata  & \nodata 
   & 423.9/434 \\ 
  \nodata  
 & \errtwo{1.077}{0.010}{0.007} 
 & \errtwo{0.252}{0.002}{0.001} 
 & \errtwo{2.78}{0.03}{0.02} 
 & \errtwo{76.1}{2.7}{2.8} 
 & \errtwo{0.3}{0.0}{0.1} 
 & \errtwo{1.1}{0.0}{0.1} 
 & \errtwo{0.6}{0.2}{0.2} 
 & \errtwo{-6.6}{0.2}{0.3} 
   & \nodata 
   & 1437.0/732 \\ 
 8 
 & \errtwo{1.422}{0.013}{0.026} 
 & \errtwo{0.269}{0.012}{0.004} 
 & \errtwo{2.79}{0.04}{0.03} 
 & \errtwo{75.1}{4.1}{4.8} 
 & \errtwo{0.8}{0.1}{0.4} 
 & \errtwo{4.3}{0.3}{0.7} 
 & \errtwo{0.6}{0.3}{0.2} 
  & \nodata  & \nodata 
   & 386.2/400 \\ 
  \nodata  
 & \errtwo{1.043}{0.010}{0.008} 
 & \errtwo{0.290}{0.002}{0.002} 
 & \errtwo{2.85}{0.04}{0.03} 
 & \errtwo{82.9}{4.1}{3.3} 
 & \errtwo{0.3}{0.0}{0.0} 
 & \errtwo{1.1}{0.0}{0.1} 
 & \errtwo{0.6}{0.3}{0.2} 
 & \errtwo{-8.7}{0.3}{0.3} 
   & \nodata 
   & 885.5/698 \\ 
 9 
 & \errtwo{1.502}{0.011}{0.011} 
 & \errtwo{0.227}{0.003}{0.003} 
 & \errtwo{3.25}{0.10}{0.12} 
 & \errtwo{1.8}{2.1}{1.5} 
 & \errtwo{7.2}{3.7}{3.8} 
 & \errtwo{1.3}{0.4}{0.9} 
 & \errtwo{0.0}{0.2}{0.2} 
  & \nodata  & \nodata 
   & 319.2/280 \\ 
  \nodata  
 & \errtwo{1.184}{0.008}{0.006} 
 & \errtwo{0.224}{0.002}{0.000} 
 & \errtwo{2.71}{0.07}{0.08} 
 & \errtwo{2.8}{0.8}{0.7} 
 & \errtwo{0.2}{0.0}{0.0} 
 & \errtwo{0.8}{0.1}{0.0} 
 & \errtwo{0.0}{0.2}{0.1} 
 & \errtwo{-8.1}{0.3}{0.3} 
   & \nodata 
   & 917.9/578 \\ 
 10 
 & \errtwo{1.510}{0.009}{0.010} 
 & \errtwo{0.151}{0.002}{0.002} 
 & \errtwo{3.39}{0.01}{0.06} 
 & \errtwo{0.0}{0.7}{0.0} 
 & \errtwo{2.4}{0.2}{1.4} 
 & \errtwo{2.0}{0.3}{1.7} 
 & \errtwo{-0.9}{0.3}{0.3} 
  & \nodata  & \nodata 
   & 276.5/224 \\ 
  \nodata  
 & \errtwo{1.191}{0.006}{0.006} 
 & \errtwo{0.147}{0.001}{0.000} 
 & \errtwo{2.90}{0.08}{0.05} 
 & \errtwo{0.0}{0.3}{0.0} 
 & \errtwo{0.2}{0.0}{0.0} 
 & \errtwo{1.1}{0.1}{0.1} 
 & \errtwo{-0.9}{0.3}{0.2} 
 & \errtwo{-6.1}{0.4}{0.4} 
 & \errtwo{-8.0}{0.4}{0.4} 
   & 1005.1/816 \\ 
\enddata 
\tablecomments{\tcompkerrindi} \label{tab:polykerr_indi} 
\end{deluxetable*}